\documentclass[twocolumn]{aastex61}
\usepackage{multirow}
\usepackage{color}

\accepted{November 30, 2018}

\shorttitle{Tsuge et al.}
\shortauthors{Tsuge et al.}

\begin{document}

\title{Formation of the active star forming region LHA 120-N 44 triggered by tidally-driven colliding H{\sc i} flows}



\author{Kisetsu Tsuge}
\affil{Department of Physics, Nagoya University, Furo-cho, Chikusa-ku, Nagoya 464-8601, Japan; tsuge@a.phys.nagoya-u.ac.jp}

\author{Hidetoshi Sano}
\affil{Department of Physics, Nagoya University, Furo-cho, Chikusa-ku, Nagoya 464-8601, Japan; tsuge@a.phys.nagoya-u.ac.jp}
\affil{Institute for Advanced Research, Nagoya University, Furo-cho, Chikusa-ku, Nagoya 464-8601, Japan}


\author{Kengo Tachihara}
\affil{Department of Physics, Nagoya University, Furo-cho, Chikusa-ku, Nagoya 464-8601, Japan; tsuge@a.phys.nagoya-u.ac.jp}

\author{Cameron Yozin}
\affiliation{ICRAR, M468, The University of Western Australia, 35 Stirling Highway, Crawley Western Australia 6009, Australia}

\author{Kenji Bekki}
\affiliation{ICRAR, M468, The University of Western Australia, 35 Stirling Highway, Crawley Western Australia 6009, Australia}

\author{Tsuyoshi Inoue}
\affil{Department of Physics, Nagoya University, Furo-cho, Chikusa-ku, Nagoya 464-8601, Japan; tsuge@a.phys.nagoya-u.ac.jp}

\author{Norikazu Mizuno}
\affiliation{National Astronomical Observatory of Japan, Mitaka, Tokyo 181-8588, Japan}

\author{Akiko Kawamura}
\affiliation{National Astronomical Observatory of Japan, Mitaka, Tokyo 181-8588, Japan}

\author{Toshikazu Onishi}
\affiliation{Department of Physical Science, Graduate School of Science, Osaka Prefecture University, 1-1 Gakuen-cho, Naka-ku, Sakai, Osaka 599-8531, Japan}

\author{Yasuo Fukui}
\affil{Department of Physics, Nagoya University, Furo-cho, Chikusa-ku, Nagoya 464-8601, Japan; tsuge@a.phys.nagoya-u.ac.jp}
\affil{Institute for Advanced Research, Nagoya University, Furo-cho, Chikusa-ku, Nagoya 464-8601, Japan}


\begin{abstract}
N44 is the second active site of high mass star formation next to R136 in the Large Magellanic Cloud (LMC). We carried out a detailed analysis of H{\sc i} at 60 arcsec resolution by using the ATCA  \& Parkes data. We presented decomposition of the H{\sc i} emission into two velocity components (the L- and D-components) with the velocity separation of $\sim$60 km s$^{-1}$. In addition, we newly defined the I-component whose velocity is intermediate between the L- and D-components. The D-component was used to derive the rotation curve of the LMC disk, which is consistent with the stellar rotation curve \citep{2000ApJ...542..789A}. Toward the active cluster forming region of LHA 120-N 44, the three velocity components of H{\sc i} gas show signatures of dynamical interaction including bridges and  complementary spatial distributions. We hypothesize that the L- and D-components have been colliding with each other since 5 Myrs ago and the interaction triggered formation of the O and early B stars ionizing N44. In the hypothesis the I-component is interpreted as decelerated gas in terms of momentum exchange in the collisional interaction of the L- and D-components. In the N44 region the $Planck$ sub-mm dust optical depth is correlated with the H{\sc i} intensity, which is well approximated by a linear regression. We found that the N44 region shows a significantly steeper regression line than in the Bar region, indicating less dust abundance in the N44 region, which is ascribed to the tidal interaction between the LMC with the SMC 0.2 Gyrs ago.

\end{abstract}


\section{Introduction}\label{sec:intro}
\subsection{High-mass star formation}\label{sec:intro1}
The formation mechanism of high-mass stars is one of the most important issues in astronomy, because their extremely energetic interactions with surrounding material, in the form of UV radiation, stellar winds, and supernova explosions, are influential in driving galaxy evolution. Two models, the competitive accretion and the monolithic collapse have been major theoretical schemes of high-mass star formation \citep[see for reviews,][]{2007ARA&A..45..481Z,2014prpl.conf..149T}. In spite of the numerous studies, we have not understood detailed mechanisms of high-mass star formation.

Recently, the cloud-cloud collision model \citep[the CCC model;][]{1992PASJ...44..203H,2010MNRAS.405.1431A} attracted attention as a mechanism of high-mass star formation and observational studies showed that more than 30 regions of high-mass stars and/or clusters are triggered by cloud-cloud collision: NGC 1333 \citep{1976ApJ...209..466L}, Sgr B2 \citep{1994ApJ...429L..77H,2000ApJ...535..857S}, Westerlund 2 \citep{2009ApJ...696L.115F,2010ApJ...709..975O}, NGC 3603 \citep{2014ApJ...780...36F}, RCW 38 \citep{2016ApJ...820...26F}, M 42 \citep{2018ApJ...859..166F}, NGC 6334/NGC 6357 \citep{2018PASJ...70S..60F}, M17 \citep{2018PASJ...70S..42N}, W49 A \citep{1986ApJ...305..353M,2009PASJ...61...39M}, W51 \citep{2001PASJ...53..793O,2017arXiv171101695F}, W33 \citep{2018PASJ...70S..50K}, M20 \citep{2011ApJ...738...46T,2017ApJ...835..142T}, RCW 120 \citep{2015ApJ...806....7T}, [CPA2006] N37 \citep{2016ApJ...833...85B}, GM 24 \citep{2018PASJ...70S..44F}, M16 \citep{2017arXiv170606002N}, RCW 34 \citep{2018PASJ...70S..48H}, RCW 36 \citep{2018PASJ...70S..43S}, RCW 32 \citep{2018PASJ...70S..49E}, RCW 166 \citep{2018PASJ...70S..47O}, Sh2-48 \citep{2017arXiv170607164T}, Sh2-252 \citep{2013ApJ...768...72S}, Sh2-235, 237, 53 \citep{2016ApJ...819...66D,2017ApJ...834...22D,2018ApJ...861...19D}, LDN 1004E in the Cygnus OB 7 \citep{2014ApJ...797...58D},  {[CPA2006] S 36 \citep{2017ApJ...840..111T}}, [CPA2006] N35 \citep{2018PASJ...70S..51T}, [CPA2006] N4 \citep{2018arXiv180800709F}, [CPA2006] S44 \citep{2018arXiv180900118K}, {[CPA2006] N36 \citep{2018ApJ...866...20D}, [CPA2006] N49 \citep{2017ApJ...851..140D}}, LHA 120-N159 West \citep{2015ApJ...807L...4F},  [CPA2006] S 116 \citep{2018PASJ...70S..46F},  LDN 1188 \citep{2017ApJ...835L..14G},  LDN 1641-N \citep{2012ApJ...746...25N}, Serpens South \citep{2014ApJ...791L..23N},  NGC 2024 \citep{2017arXiv170605652O},  RCW 79 \citep{2018PASJ...70S..45O}, LHA 120-N159 East \citep{2017ApJ...835..108S}, NGC 2359 \citep{2017arXiv170808149S},  Circinus-E cloud \citep{2011ApJ...731...23S}, G337.916 \citep{2017ApJ...840..111T}, {Rosette Molecular Cloud \citep{2018ApJS..238...10L}}, Galactic center \citep{2015PASJ...67..109T}, {G35.2N \& G35.2S \citep{2017ApJ...837...44D}}, NGC 2068/NGC 2071 \citep{2017arXiv170605664T}, NGC 604 \citep{2018PASJ...70S..52T} These studies lend support for the role of cloud-cloud collision which triggers formation of O stars. A typical collision size scale and relative velocity between the two clouds are 1--10 pc and 10--20 km s$^{-1}$ and the number of O-type stars formed by the collision is in a range from 1 to $\sim$20 {for H$_{2}$ column density greater than 10$^{22}$ cm$^{-2}$. The CCC model realizes gas compression by two orders of magnitude within 10$^{5}$ yrs ($\sim$1 pc/10 km s$^{-1}$)} and provides a high mass accretion rate by supersonic collision between molecular clouds. These phenomena clearly creates physical conditions which favor high-mass star formation according to theoretical simulations of the collisional process in the realistic inhomogeneous molecular gas \citep{2013ApJ...774L..31I,2018PASJ...70S..53I}, whereas it remains to be observationally established how unique and common the role of the collisions is in the O-type star formation. In this context we note that a large-scale model of molecular cloud evolution via collision demonstrates the important role of collision in star formation and cloud growth \citep{2018PASJ...70S..59K}.

One of the observational signatures of cloud-cloud collisions is complementary distribution between the two colliding clouds. It is usual that the colliding two clouds have different sizes as simulated by theoretical studies \citep{1992PASJ...44..203H,2010MNRAS.405.1431A,2014ApJ...792...63T}. If one of the colliding clouds is smaller than the other, the small cloud creates a cavity of its size in the large cloud through the collision. The cavity is observed as an intensity depression in the gas distribution at a velocity range of the large cloud, and the distributions of the small cloud and the cavity exhibit a complementary spatial distribution. The complementary distributions often have some displacement because the relative motion of the collision generally makes a non-negligible angle to the line of sight. \cite{2018ApJ...859..166F} presented synthetic observations by using the hydrodynamical numerical simulations of \cite{2014ApJ...792...63T}. {In the previous works, there are more than 6 regions where the displacement was well determined, including M17 \citep{2018PASJ...70S..42N}, M42 \citep{2018ApJ...859..166F}, RCW 36 \citep{2018PASJ...70S..43S}, GM 24 \citep{2018PASJ...70S..44F}, S 116 \citep{2018PASJ...70S..46F}, R136 (Pape I).}



{The collisional process on a few--10 pc scale} were investigated by Magnetohydrodynamic (MHD) numerical simulations. These simulations show that the H{\sc i} gas flow colliding at 20 km s$^{-1}$ is able to compress H{\sc i} gas and the density of H{\sc i} gas is enhanced in the compressed layer \citep{2012ApJ...759...35I}. Molecular clouds are formed in the shock-compressed layer in a time scale of $\sim$10 Myr, which can probably become shorter to a few Myr when the colliding velocity is as fast as 100 km s$^{-1}$ if density is higher. These clouds will become self-gravitating when they grow massive enough. \cite{2013ApJ...774L..31I} and \cite{2018PASJ...70S..53I} calculated the subsequent physical process where two molecular gas flows collides at 20 km s$^{-1}$. These studies show that strong shock waves generated by a cloud-cloud collision form gravitationally unstable massive molecular cores, each of which directly leads to form a high-mass star.

\subsection{Previous study of the LMC}\label{sec:intro2}

In the context of cloud-cloud collision, the nearest galaxies to the Galaxy, the LMC and the SMC, provide an excellent laboratory to test star formation and the collisional triggering. \cite{1992A&A...263...41L} separated two H{\sc i} velocity components of the LMC whose velocity difference is $\sim$50 km s$^{-1}$ ; one is the H{\sc i} gas extending over the whole disk of the LMC (hereafter D-component) and the other more spatially confined H{\sc i} having lower radial velocity (hereafter L-component). They analyzed the H{\sc i} data obtained by Parkes telescope with angular resolution and grid size of 15\arcmin  and 12\arcmin, respectively, and they found that the D- and L-components contain 72 \% and 19 \% of the whole H{\sc i} gas, respectively. In spite of the work by \cite{1992A&A...263...41L}, the implications of the two components on star formation was not observationally explored for nearly two decades.

\begin{figure*}[htbp]
\begin{center}
\includegraphics[width=\linewidth]{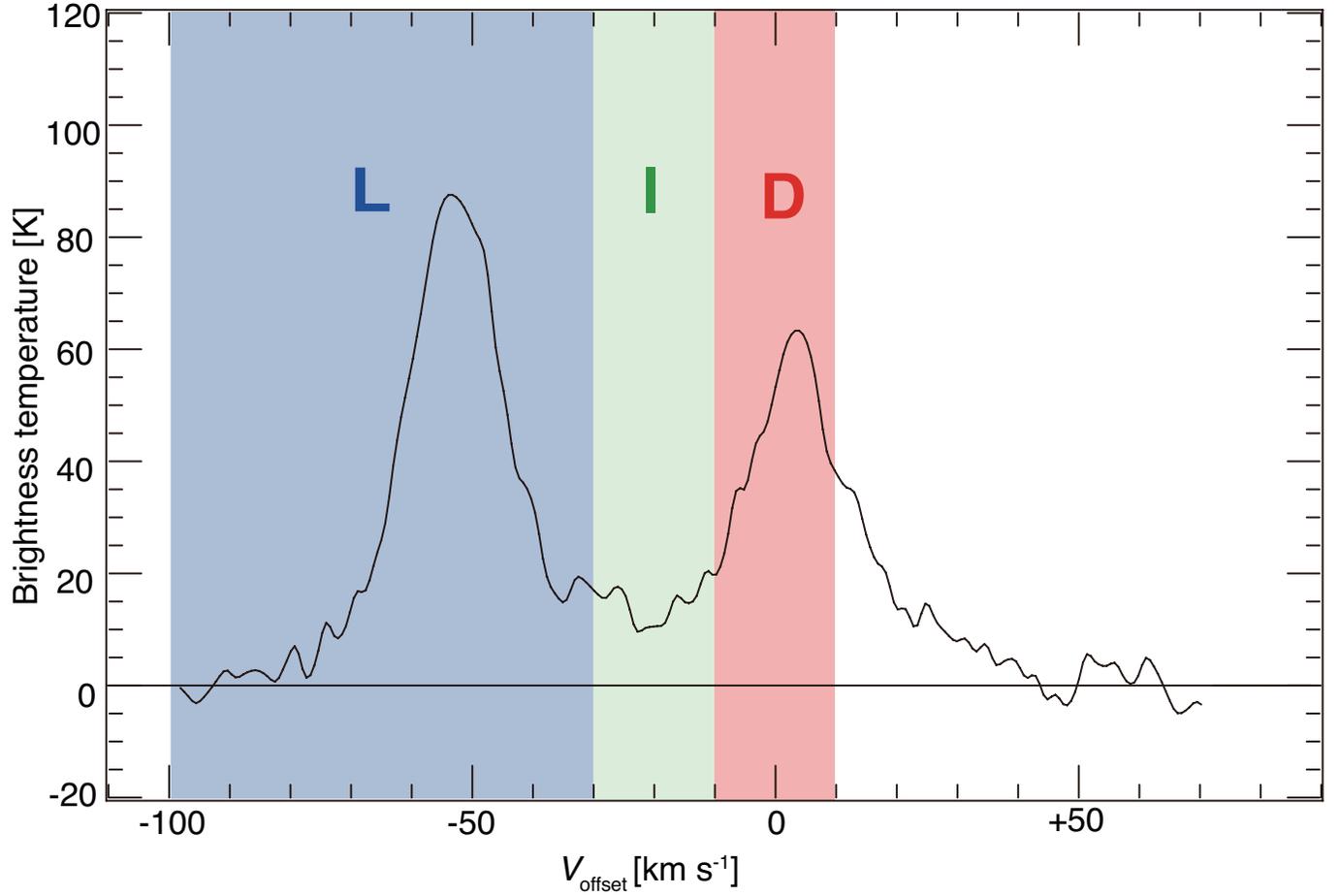}
\end{center}
\caption{The typical spectra of H{\sc i} at (R.A., Dec.)=(5$^{h}$46$^{m}$43.54$^{s}$, -69$^{d}$42$^{m}$59.31$^{s}$).} 
\label{fig0-1}
\end{figure*}

Recently it was found that the young massive cluster RMC136 (R136) in the Large Magellanic Cloud (LMC) is located toward an overlapping area of the L- and D-components \citep[][;hereafter Paper I]{2017PASJ...69L...5F}. These authors analyzed the H{\sc i} data of the whole LMC at a 1\arcmin \ resolution which corresponds to 15 pc at a distance of the LMC \citep{2003ApJS..148..473K}, {and separated the L- and D-components at a higher resolution. Figure \ref{fig0-1} shows a typical H{\sc i} spectrum which presents the L- and D-components. The authors of paper I revealed} that the two components are linked by the bridge features in the velocity space and show complementary spatial distributions on a kpc scale. Based on these two signatures it was suggested that the L- and D-components are colliding toward R136, and a scenario was presented that the collision of the H{\sc i} flows triggered the formation of the R136 cluster and high mass stars in its surroundings. They also showed that a component having intermediate velocity between the L- and D-components (hereafter I-component), and indicated that the I-component may represents merging of the L- and D-components where their relative velocity is decelerated with momentum conservation (see Figure 2 of Paper I). According to the theoretical studies including hydrodynamical numerical simulations, the LMC and SMC had a close encounter 0.2 Gyr ago and their tidal interaction stripped gas from the two galaxies. Currently the remnant gas is falling down to each galaxy, and observed as the H{\sc i} gas with significant relative velocity \citep{1990PASJ...42..505F,2007PASA...24...21B,2014MNRAS.443..522Y}.

\subsection{LHA 120-N 44}\label{sec:intro3}
We extend the study from R136 to other active H{\sc ii} regions in the LMC and to test how the colliding gas flow leads to massive star formation. Following Paper I, we analyze the region of LHA 120-N 44 (N44) in the present paper. N44 is the H{\sc ii} region cataloged by \cite{1956ApJS....2..315H} and is one of the most active star-forming regions in the LMC. N44 is older \citep[5--6 Myr: ][]{1997A&AS..123..455W} than R136 \citep[1.5--4.7 Myr:][]{2018Sci...359...69S}. N44 holds the second largest number of O / WR stars, $\sim$40, corresponding to $\sim$1/10 of R136 \citep{2009AJ....138.1003B}, in the LMC, and N44 has received much attention at various wavelengths from mm/sub-mm, infrared, optical to high-energy X-rays \citep[e.g.,][]{1998ApJ...503..729K,2009ApJ...695..511C,2012A&A...542A..66C}. This region is also categorized as a superbubble by a H$\alpha$ shell, and the OB association LH 47 is located in the center of the shell. \cite{2016MNRAS.457.2048A} investigated kinematic features of H$\alpha$ emission and compared it with the L- and D-components. These authors showed that compact H{\sc ii} regions N44B and C have two H$\alpha$ components with a velocity difference of $\sim$30 km s$^{-1}$, and interpreted that N44B and C belong to the L- and D-components, respectively. \cite{1998ApJ...503..729K} analyzed the high-resolution H{\sc i} data toward N44 and indicated that the H{\sc i} shell corresponding to the H$\alpha$ shell was produced by stellar winds and supernovae, whereas the relationship between the shells and the L- and D-components remain unknown. These previous studies show that the N44 region is the most suitable target for studying high mass star formation next to R136.

The present paper is organized as follows; Section 2 summarizes the data sets and methodology and Section 3 the results. The discussion is given in Section 4, in which we test whether the high-mass star formation in N44 have been triggered by the colliding H{\sc i} flows. Section 5 gives a summary.

\section{Data set and masking}\label{sec:data}
\subsection{Data sets}
\subsubsection{\rm{H\sc{i}}}

Archival data of the Australia Telescope Compact Array (ATCA) and Parkes H{\sc i} 21 cm line emission are used \citep{2003ApJS..148..473K}. The angular resolution of the combined H{\sc i} data is 60$\arcsec$ ($\sim$15 pc at the LMC). The rms noise level of the data is $\sim$2.4 K for a velocity resolution of 1.649 km s$^{-1}$. They combined the H{\sc i} data obtained by ATCA \citep{1998ApJ...503..674K} with the these obtained with the Parkes multibeam receiver with resolution of 14$\arcmin$--16$\arcmin$ \citep{1997PASA...14..111S}. This is because the ATCA is not sensitive to structures extending over 10$\arcmin$--20$\arcmin$ (150--160 pc) due to the minimum baseline of 30 m.

\subsubsection{$Planck$/$IRAS$}
Archival datasets of the dust optical depth at 353 GHz ($\tau_{353}$) and dust temperature ($T_{\rm d}$) were obtained by using the combined $Planck$/$IRAS$  data with the gray-body fitting. For details, see \cite{2014A&A...571A..11P}. These datasets are used to make comparison with the H{\sc i} data. The angular resolution is 5$\farcm$0 ($\sim$ 75 pc at the LMC) with a grid spacing of 2$\farcm$4. For comparison of $\tau_{353}$ with H{\sc i}, the spatial resolution and the grid size of H{\sc i} were adjusted to $\tau_{353}$. 

\subsubsection{\rm{CO}}

We used $^{12}$CO($J$ = 1--0) data of the Magellanic Mopra Assessment \citep[MAGMA;][]{2011ApJS..197...16W} for a small-scale analysis in the N44 region. The effective spatial resolution is 45$\arcsec$ (11 pc at a distance of 50 kpc), and the velocity resolution is 0.526km s$^{-1}$. The MAGMA survey does not cover the whole LMC and the observed area is limited toward the individual CO clouds detected by NANTEN \citep{1999PASJ...51..745F,2001PASJ...53..971M}.

We used the data of the $^{12}$CO($J$ = 1--0) observed over the whole LMC with the NANTEN 4-m telescope at a resolution of 2$\farcm$6, a grid spacing of 2$\farcm$0, and velocity resolution of 0.65 km s$^{-1}$ \citep{1999PASJ...51..745F,2008ApJS..178...56F,2009gcgg.book..121K}. In the present paper, the NANTEN data were used to mask the CO emitting regions in comparing the $Planck$/$IRAS$ data with the H{\sc i}.

\subsubsection{H$\alpha$}

H$\alpha$  data of the Magellanic Cloud Emission-Line Survey \citep[MCEL;][]{1999IAUS..190...28S} are used for a comparison of spatial distribution with H{\sc i}. The dataset was obtained with a 2048 $\times$ 2048 CCD camera on the Curtis Schmidt Telescope at Cerro Tololo Inter-American Observatory. The angular resolution is $\sim$3\arcsec--4\arcsec ($\sim$0.75--1.0 pc at distance of 50 kpc). We also use the archival data of H$\alpha$ provided by the Southern H--Alpha Sky Survey Atlas \citep[SHASSA; ][]{2001PASP..113.1326G} in order to define the region where UV radiation is locally enhanced. The same method was also applied in Paper I, and hence we can compare the result of R136 with that of N44. The angular resolution is about 0$\farcm$8, and the sensitivity level is 2 Rayleigh (1.2$\times$10$^{-17}$ erg cm$^{-2}$ s$^{-1}$ arcsec$^{-2}$) pixel$^{-1}$.

\subsection{Masking}
We masked some regions in the distribution of the dust optical depth at 353 GHz ($\tau_{353}$ ) in order to avoid the effect of local dust destruction by the ultraviolet (UV) radiation from O-type stars/H{\sc ii} regions in comparison of $\tau_{353}$ with H{\sc i}. Such a local effect may alter the dust properties which is assumed to be uniform in the comparison, and can distort the correlation between H{\sc i} and dust emission. By using CO and H$\alpha$ data, the regions with CO emission higher than 1 K km s$^{-1}$ ($\sim$1$\sigma$) with H$\alpha$ emission higher than 30 Rayleigh are masked as made in the previous works \citep{2014ApJ...796...59F,2015ApJ...798....6F,2017ApJ...838..132O}.

\section{Results}\label{sec:results}
\subsection{Method of H{\sc i} spectral decomposition}
In Paper I H{\sc i} gas decomposition into the L- and D-components were demonstrated, but details were not given. We here describe a detailed method of the decomposition used in the present paper and in Paper I in the Appendix. In order to decompose the H{\sc i} spectra into the L- and D-components, we made Gaussian fittings to the H{\sc i} data. We fitted the Gaussian function to the H{\sc i} emission lines for each pixel with peak intensity greater than 20 K and derived the peak velocity of the spectrum. This process allowed us to derive the velocity of the D-component toward the areas where the L-component is weaker than the D-component. For the profiles showing the L-component as the primary peak we made another fitting to the secondary peak and define the D-component if the secondary peak velocity is larger than the primary peak. We thus obtained distribution of the D-component over the disk. The result was used to derive the rotation velocity of the disk. Then, we subtracted the rotation velocity for each pixel by shifting the spectra in the velocity channel over the LMC. We defined the $V_{\rm offset}$ as the relative velocity to the D-component as follows: $V_{\rm offset}$ = $V_{\rm LSR}$$-$$V_{\rm D}$ ($V_{\rm D}$ = the radial velocity of the D-component). Hereafter, we use $V_{\rm offset}$ instead of $V_{\rm LSR}$ in the following.

We determined the typical velocity ranges of the L- and D-components as $V_{\rm offset}$ from $-$100 to $-$30 km s$^{-1}$ and from $-$10 to +10 km s$^{-1}$, respectively, over the whole galaxy from position-velocity diagrams, and velocity channel map in Paper I. We also defined the intermediate velocity between the L- and D-components in a range of $V_{\rm offset}$ from $-$30 to $-$10 km s$^{-1}$ (I-component).The maximum variation of velocity range is about $\pm$ 10 km s$^{-1}$.

We note that these velocity ranges vary from region to region, which require fine tuning in the individual regions due to velocity variation of the L-component.

\subsection{The rotation curve of the LMC}\label{subsec:RC}

We derived the rotation curve of the LMC represented by the velocity distribution of the D-component as shown in Figure \ref{figB1}c. As described in the Appendix B, we optimized the coordinate of rotation center, the inclination angle of the galaxy disk plane, and the position angle of the inclination axis. The present curve shows the flat rotation of the LMC with the constant rotation velocity of $\sim$60 km s$^{-1}$ at the galactocentric radius ($R_{\rm G}$) is larger than 1 kpc, and is consistent with the stellar disk rotation as shown in Figure \ref{figB1}c. The previous rotation velocity of H{\sc i} \citep{1998ApJ...503..674K} decreases at $R_{\rm G} >$ 2.5 kpc. \cite{1998ApJ...503..674K} did not consider the two velocity components and their rotation velocity is from a mixture of the L- and D-components. Moreover, it is likely that  the extended emission was resolved out since they used the H{\sc i} data taken with the ATCA alone. In the present study, we used the combined with ATCA and Parkes H{\sc i} data and the missing flux is recovered.

\subsection{Distribution of the L- and D-components}
We compare H{\sc i} distribution with the major star forming regions over the whole LMC. Figures  \ref{fig1}a and  \ref{fig1}b present distributions of the L- and D-components, and Figure  \ref{fig1}c an overlay of the two components. The L-component  distribution is concentrated in two regions. One of them is the H{\sc i} Ridge stretching by $\sim$1 kpc $\times$ 2.5 kpc in R.A. and Dec., including young star forming regions R136 and LHA 120-N 159. The other region is extended to the northwest from the center of the LMC with a size of $\sim$2 kpc $\times$ 2 kpc including LHA 120-N 44 (hereafter “diffuse L-component”). The mean intensity of the diffuse L-component is $\sim$40 \% of that of the H{\sc i} Ridge. Several H{\sc ii} regions \citep{2016MNRAS.457.2048A} are located in the south of the diffuse L-component. The D-component is extended over the entire LMC and is characterized by the morphological properties with many cavities and voids \citep[e.g.,][]{1998LNP...506..521K,2013ApJ...763...56D}.


\subsection{Velocity distribution of the H{\sc i} gas {toward the N44 region}}
Figure \ref{fig2}a shows H{\sc i} total intensity distribution toward the presently analyzed region of N44. H{\sc i} gas is concentrated in the region within $\sim$200 pc of N44C and its H{\sc i} integrated intensity is enhanced to 2000 K km s$^{-1}$. Figure \ref{fig2}b is a H$\alpha$ image \citep[MCEL;][] {1999IAUS..190...28S} of the star forming region indicated by a black box in Figure \ref{fig2}a. N44 B and N44 C are located in a shell (white dashed circle) whose central coordinate is (R.A., Dec.)$\sim$(5$^{h}$22$^{m}$15$^{s}$, -67$^{d}$57$^{m}$00$^{s}$). The shell includes a WR-star and 35 O-type stars. In the southeast, there is another H{\sc ii} region N44D containing two O-type stars \citep{2009AJ....138.1003B}.

Figure \ref{fig3} shows typical spectra of H{\sc i} in the northeast and south of N44. There are two velocity components with $V_{\rm offset}$ of $\sim$$-$60 km s$^{-1}$ and $\sim$$-$30 km s$^{-1}$, besides the D-component. In Figure \ref{fig3}a, we find the L-component is peaked at $-$60 km s$^{-1}$ , 10 km s$^{-1}$ lower than that shown in Figure \ref{fig0-1}, while the D-component is peaked at 0 km s$^{-1}$. The L- and the D-components are connected by a bridge feature between the two velocity components. We interpret the component at $V_{\rm offset}$ = $-$80.6--$-$50.0 km s$^{-1}$ as part of the “L-component”, that at $V_{\rm offset}$ =$-$50.0--$-$15.6 km s$^{-1}$ as part of the “I-component”. In Figure 5b, no L-component is seen and we interpret that there is only the I-component at $V_{\rm offset}$ =$-$50.0--$-$15.6 km s$^{-1}$. In comparing the H{\sc i} with the O-type stars we focus on Lines A, B and C in Figure \ref{fig2}a where most of the massive stars in N44 are concentrated.

Figure \ref{fig4} shows Dec.-velocity diagrams integrated in the R.A. direction along Lines A, B, and C which have a width of $\sim$87 pc and a length of $\sim$1.1 kpc in Figure \ref{fig2}a. Lines A, B, and C pass the eastern, central, and western regions of N44, respectively.
 In Line A (Figure \ref{fig4}a), the D-component and the L-component are located at Dec.=$-$67$^{d}$45$^{m}$00$^{s}$ to $-$68$^{d}$40$^{m}$00$^{s}$ and $-$67$^{d}$50$^{m}$00$^{s}$ to $-$68$^{d}$0$^{m}$00$^{s}$, respectively. The bridge feature connects them at the position shown by a white dashed box in velocity space.
In Line B (Figure \ref{fig4}b), the D-, and I -components are seen at Dec.=$-$67$^{d}$30$^{m}$00$^{s}$ to $-$68$^{d}$09$^{m}$00$^{s}$ and Dec.=$-$68$^{d}$03$^{m}$00$^{s}$ to $-$68$^{d}$15$^{m}$00$^{s}$, respectively. A bridge feature which is shown by a white dashed box connects them. The D-, I -, and L-components are distributed in Line C (Figure \ref{fig4}c). We also see bridge features like in the other Lines as indicated by white dashed boxes. At the position of Dec.=$-$68$^{d}$06$^{m}$00$^{s}$ to $-$68$^{d}$15$^{m}$00$^{s}$, the L- and D-components are connected by another bridge feature, and at Dec.=$-$68$^{d}$30$^{m}$00$^{s}$ -- $-$68$^{d}$40$^{m}$00$^{s}$, the I - and D-components are connected by a bridge feature.

\subsection{The spatial distribution of H{\sc i} gas {toward the N44 region}}
Detailed spatial distributions of the L-, I-, and D-components {around N44 are shown in Figure \ref{fig5}}. The L-component is distributed in the northeastern and southwestern regions of N44 (Figure \ref{fig5}a). The I -component is extended by $\sim$200 pc $\times$ 400 pc in the south of N44 (Figure \ref{fig5}b). {The D-component mainly spreads from the central region to southeast of N44 (Figure \ref{fig5}c).}

Comparisons of spatial distributions of different velocity components are shown in Figure \ref{fig6}, which shows the distribution of the D-component ($-$15.6--+9.7 km s$^{-1}$) overlaid on the L-component contours. The L-component has complementary distributions with the D-component at the positions of (RA., Dec.) = (5$^{h}$22$^{m}$30$^{s}$--5$^{h}$24$^{m}$0$^{s}$, $-$67$^{d}$50$^{m}$0$^{s}$--$-$68$^{d}$0$^{m}$0$^{s}$) and (5$^{h}$20$^{m}$0$^{s}$--5$^{h}$22$^{m}$30$^{s}$,$-$68$^{d}$20$^{m}$0$^{s}$--$-$68$^{d}$0$^{m}$0$^{s}$) as indicated by black arrows. Two of the L-component exist along the edge of the D-component. {We interpret that the two depressions in the D-component, which corresponds the L-components, may represent interaction between the two components. There are more intensity depressions where no corresponding L-components are seen. They are possibly due to preexistent intensity variations or created by other mechanisms like supernovae.}

In Figure \ref{fig7}, {the D-component} shows intensity depression toward the I-component at the positions of (R.A., Dec.)$\sim$ ($-$67$^{d}$58$^{m}$0$^{s}$--$-$68$^{d}$36$^{m}$00$^{s}$, 5$^{h}$18$^{m}$0$^{s}$--5$^{h}$26$^{m}$00$^{s}$) as indicated by the red box. We find some displacement of the complementary distribution between the I- and D-components, which is a {usual} signature caused by {a certain} angle {of cloud-cloud collision} to the line of sight \citep{2018ApJ...859..166F}. We calculated the displacement by using overlapping function $H$($\Delta$) in pc$^{2}$ {which shows a degree of complementarity} \citep{2018ApJ...859..166F}. We {derived} the {projected} displacement where the overlapping area of the strong I-component (intensity larger than 550 K km s$^{-1}$) and the depression of the D-component (intensity smaller than 1100 K km s$^{-1}$) becomes maximum.

In Figure \ref{fig7}b the {I}-component is surrounded by the D-component at (R.A., Dec.) = ($-$68$^{d}$4$^{m}$0$^{s}$ -- $-$68$^{d}$42$^{m}$00$^{s}$, 5$^{h}$18$^{m}$0$^{s}$--5$^{h}$26$^{m}$00$^{s}$) as indicated by the black box. They present complementary distribution with a displacement of 107 pc and a position angle of 180 deg.

Figure \ref{fig8} shows an enlarged view of the N44 region, showing complementary distributions among the D-, I-, and L-components at a $\sim$100 pc scale. The D- and I-components have complementary distribution at (R.A., Dec.)$\sim$ (5$^{h}$21$^{m}$30$^{s}$ -- 5$^{h}$24$^{m}$0$^{s}$, $-$68$^{d}$5$^{m}$0$^{s}$ -- $-$68$^{d}$10$^{m}$0$^{s}$ ) and the L- and D-components show complementary distribution at (RA., Dec.)$\sim$(5$^{h}$23$^{m}$0$^{s}$-5$^{h}$23$^{m}$36$^{s}$, $-$67$^{d}$58$^{m}$0$^{s}$-- $-$67$^{d}$52$^{m}$0$^{s}$) and (5$^{h}$20$^{m}$30$^{s}$--5$^{h}$21$^{m}$30$^{s}$, $-$68$^{d}$10$^{m}$0$^{s}$-- $-$68$^{d}$5$^{m}$0$^{s}$). At (RA., Dec.)$\sim$ (5$^{h}$22$^{m}$30$^{s}$, $-$67$^{d}$56$^{m}$00$^{s}$), an H{\sc i} cavity of a $\sim$20 pc radius {is seen}. It is suggested that the H{\sc i} cavity indicates dynamical effects of stellar winds and supernova explosions \citep{1998ApJ...503..729K}. There is only the D-component within $\sim$60 pc of the H{\sc i} cavity.

\subsection{{Physical properties and the distribution of massive stars}}

We calculate the mass and column density on an assumption that H{\sc i} emission is optically thin. We use the equation as follows \citep{1990ARA&A..28..215D}: 

\begin{eqnarray}
N_{\rm H{\sc I}} = 1.8224\times10^{18}\int \Delta T_{\rm b}  \it dv \ \rm{[cm^{-2}] },
\end{eqnarray}

where $T_{\rm b}$ is the observed H{\sc i} brightness temperature [K].

The physical properties of H{\sc i} gas along Lines A, B, and C (width: $\sim$70 pc, length: $\sim$360 pc) shown in Figure \ref{fig8} are summarized in Table 1. We integrated all the velocity range ($-$100.1 $<$ $V_{\rm offset}$ $<$ 89.7 km s$^{-1}$) for estimating mass and column density of H{\sc i} gas. The mass and column density of Lines A, B, and C show similar values. The mass is about 10$^{6}$ $M_{\sun}$ and the peak column density is about 6$\times$10$^{21}$ cm$^{-2}$. The velocity separations between the two velocity components in Lines A, B, and C are about 20 to 60 km s$^{-1}$. The bridge features are observed in {Lines A, B, and C}. Most of the massive stars ({WR and O-type stars}) are located in Line B.
{Figure \ref{fig9} shows a histogram of the number of massive stars around the N44 region in Figure \ref{fig8}. We counted the number of WR and O-type stars located inside the areas of Lines A, B, and C  \citep[e.g.,][]{2009AJ....138.1003B}. Lines A, B, and C have 2, 32, and 5 massive stars, respectively.}

We calculate the mass of the L-, I -, and D-components toward {the} N44 region in Figure \ref{fig8}. Table 2 summarizes the physical quantities around N44. We used equation (1) for the calculation of H{\sc i} {parameters} in the same method as in Table 1. The total mass is  $\sim$ 3.0 $\times$ 10$^{6}$ $M_{\odot}$ {and} the mass of the D-component occupies 2/3 of the total mass.

H$_{2}$ gas {mass was estimated} by using the $W_{\rm CO}$--$N$(\rm H$_{2}$) conversion factor \citep[$X_{\rm CO}$= 7.0$\times$10$^{20}$ cm$^{-2}$ (K km s$^{-1}$)$^{-1}$;][]{2008ApJS..178...56F}. We use the equation as follows: 
\begin{eqnarray}
N (\rm H_{2}) = \it X_{\rm CO} \times \it{W} (^{\rm 12}\rm{CO}(\it{J}\rm{ = 1-0})),
\end{eqnarray}
where $W_{\rm{CO}}$ is the {integrated} intensity of the $^{\rm 12}\rm{CO}(\it{J}\rm{ = 1-0})$ and $N$(\rm H$_{2}$) is the column density of molecular hydrogen. The total mass of H$_{2}$ gas is $\sim$ 3.0$\times$10$^{6}$ $M_{\sun}$. {Most} of CO clouds {are included} in the D-component.


\subsection{Comparison with dust emission}
We compared the intensity of H{\sc i} ($W$$_{\rm H{\sc I}}$) and {the dust} optical depth at 353 GHz ($\tau_{353}$) in order to investigate {the gas to dust ratio}. Figure \ref{fig10} shows a typical spectrum of H{\sc i} \citep{2016A&A...594A.116H} toward the LMC. There is a Galactic foreground component at $-$50 $<$ $V_{\rm LSR}$ $<$ + 50 km s$^{-1}$, while H{\sc i} emission in +150 $<$ $V_{\rm LSR}$ $<$ + 350 km s$^{-1}$  belongs to the LMC. Because $\tau_{353}$ does not have velocity information, we subtracted the foreground dust {optical depth} from $\tau_{353}$ by assuming that the foreground H{\sc i} emission is optically thin and its $W$(H{\sc i}) is proportional to $\tau_{353}$ using $W$$_{\rm H{\sc I}}$-$\tau_{353}$ relation in the Galaxy \citep{2015ApJ...798....6F} as follows:


\begin{eqnarray}
W_{\rm H{\sc I}}  [\rm {K \ km s^{-1}}] = 1.15\times10^{8} [\rm {K \ km s^{-1}}] \times \it \tau_{\rm 353}
\end{eqnarray}

{It is also possible to subtract the Galactic foreground by} {using an uniform Galactic foreground contamination using the averaged $\tau_{353}$ value around the LMC.} {However, it is not appropriate to use the data around the LMC, because H{\sc i} gas is faint but widely extend around the LMC like as the Magellanic Bridge and the Magellanic Stream \citep{1998Natur.394..752P}. On the other hand, as shown in Figure \ref{fig10}, it is better to estimate the Galactic foreground from H{\sc i} data since H{\sc i} gas can be clearly separated into the foreground component and the LMC component by using velocity information. Figures \ref{tau}a and \ref{tau}c show the spatial distributions of $\tau_{353}$ before and after subtracting the Galactic foreground, respectively. In addition, the Galactic foreground is considerably weaker than the LMC component, so the subtraction is made fairly accurately. The Galactic foreground accounts for 30\% or less of the total $\tau_{353}$ in the main regions of the LMC (Figure \ref{tau}b, d). }

First, we {smooth the spatial resolution of H{\sc i} to the same resolution as $\tau_{353}$, 5$\arcmin$, and subtract the foreground from the total $\tau_{353}$ for each pixel.} We use only the LMC component for a comparison between the H{\sc i} data and $\tau_{353}$ after the subtraction of the foreground component. Figure \ref{fig11} shows a scatter plot between $W_{\rm H{\sc I}}$ and $\tau_{353}$ for each pixel. The gray diamonds and the red circles indicate data points of the Bar region (Paper I) and that of the N44 region (Figure \ref{fig2}a), respectively. The values of $\tau_{353}$ in the LMC are small ($<$10$^{-4}$) {and the dust} emission is optically thin. Therefore, $\tau_{353}$ reflects the column density of hydrogen atoms and is proportional to $W_{\rm H{\sc I}}$ under the assumptions of (1) H{\sc i} emission is optically thin, (2) dust mixes well with gas, and (3) dust optical properties are uniform. On the other hand, saturation of $W_{\rm H{\sc I}}$ is caused when the atomic gas becomes colder and denser. It is thought that H{\sc i} is optically thin at high temperature because the absorption coefficient of the H{\sc i} emission is inversely proportional to the spin temperature. We use H{\sc i} data which keep a proportional relation between $W_{\rm H{\sc I}}$ and $\tau_{353}$. In this paper, we assumed that H{\sc i} gas is optically thin for the data points of $T_{d}$ is high (22--24.5 K). Dust temperature ($T_{d}$) varies from region to region. So, we used H{\sc i} gas at $T_{d}$$>$ 22.5 K in the Bar region, and used H{\sc i} gas data for $T_{d}$$>$24.5 K in the N44 region.

The ratio of $W$(H{\sc i}) / $\tau_{353}$ is an indicator of {the} gas-dust ratio under the above assumptions of (1)--(3). We derived the slope between $W$(H{\sc i}) and $\tau_{353}$ by the least-squares fit which is assumed to have a zero intercept. The black and red linear lines indicate the fitting results of the {Bar} region and that of the N44 region, respectively. The slope of the {Bar} region is 7.6$\times$10$^{7}$ K km s$^{-1}$ and that of N44 region is 1.0$\times$10$^{8}$ K km s$^{-1}$ showing difference of slopes by a factor of 1.3. This {implies} that the metallicity in the N44 region is lower than that in the Bar region, a trend similar to the H{\sc i} ridge (Paper I). {This result is discussed in Section {4.3}}. {The dispersion of the Bar region looks larger than the N44 region. This means that there may be metallicity gradient in the Bar region. We will present the detailed gas to dust ratio of the whole LMC in following paper (K. Tsuge et al. in preparation).}
 
\section{Discussion}\label{sec:discussion}
\subsection{Evidence for the collision of H{\sc i} {between the two velocity components}}
We revealed the spatial and velocity distributions of the L- ($V_{\rm offset}$ =$-$80.6--$-$50.0. km s$^{-1}$), I- ($V_{\rm offset}$ = $-$50.0--$-$15.6 km s$^{-1}$), and D- ($V_{\rm offset}$ =$-$15.6--+14.9 km s$^{-1}$) components by using high-resolution H{\sc i} data. These results show the {signature}s of collision toward N44 between the H{\sc i} flows {in a similar fashion} for the H{\sc i} Ridge and R136 in Paper I.

In Paper I, it was shown that the complementary spatial distribution between the L- ($V_{\rm offset}$ = $-$100.1--$-$30.5 km s$^{-1}$) and the D-components ($V_{\rm offset}$ = $-$10.4--+9.7 km s$^{-1}$) at a kpc scale. It was also noted that the L- and D-components are connected by the bridge features in velocity space. These results {give evidence of} collision between the L- and D-components, suggesting that the collision of the H{\sc i} gas triggered the formation of $\sim$400 O /WR stars in the H{\sc i} Ridge which includes R136, N159, and the other active star forming {giant molecular clouds} (Paper I). The observational signatures of high-mass star formation by colliding H{\sc i} flows are characterized by the three elements: (1) the two velocity components {with} a supersonic velocity separation, (2) the bridge features which connect the two velocity components in velocity space, and (3) the complementary spatial distribution between {the two} H{\sc i} flows. In order to test whether the high-mass star formation in N44 have been triggered by the colliding H{\sc i} flows, we compared {the} above signatures with the observational results for N44.

\subsubsection{{The supersonic velocity separation}}\label{sec:1}

{In N44, the velocity separation of the two components is about 30 to 60 km s$^{-1}$ (Figure \ref{fig3}). If we consider a projection effect, the actual velocity separation is larger than observed. These large velocity separations cannot be explained by the stellar winds and UV radiation from the high-mass stars, where the typical stellar feedback energy is less than 10\% of the kinematic energy of the colliding flows (Paper I). As we mentioned in section \ref{sec:intro3}, \cite{1998ApJ...503..729K} found that there are two clear intensity depressions of H{\sc i} gas and expanding shells of H{\sc i} gas by small-scale (hundreds pc) analysis (shown in Figures 2 and 4 of their paper). However, we did not find obvious expansion structures in a kpc scale analysis (Figure \ref{fig4}). Figure \ref{mom} shows the second moment map of the D-component ($-$15.6$<$$V_{\rm offset}$$<$+14.9 km s$^{-1}$). The average velocity dispersion is $\sim$7.6 km s$^{-1}$ and there is no spatial correlation between the H$\alpha$ shell (dashed circle) and velocity dispersion (Figure \ref{mom}a). Moreover, it is impossible to explain the motion of the L-, I-, and D-components by expanding shell from the compact area of energy supply. We calculated the kinematic energy of the H{\sc i} gas as $\sim$10$^{53}$ erg by assuming the expansion velocity is 30 km s$^{-1}$. So, the kinematic energy of a super bubble is insufficient to explain the velocity distribution of the H{\sc i} gas. Moreover, it requires $\sim$2000 supernova explosions (SNes) within the age of N44 if we assume that kinematic energy injection by SNe is transferred in gas acceleration by $\sim$5\% \citep{2012MNRAS.426.3008K}. This number of SNe is much larger than $\sim$300 which is expected from the scale of the super bubble N44 \citep{1991A&A...245..635M}.}

{On the other hand, it is seen that the velocity dispersion of the D-component increases toward the region where the I-component is located as shown in Figure \ref{mom}b. This suggests that the collision between H{\sc i} flows played a role in creating the observed gas motion. }


\subsubsection{The bridge features in velocity space}\label{sec:2}
As seen in Figure \ref{fig4}, there are {I-component} between the two velocity components. These signatures indicate that the two velocity components interact dynamically with each other in spite of the large velocity separation, which otherwise may suggest a large {spatial} separation {and no interaction between the two in space.}

\subsubsection{Complementary spatial distribution}\label{sec:3}
We revealed the detailed spatial distributions of the L-, I-, and D-components, and compared their spatial distributions. {In Figure \ref{fig7} the complementary spatial distribution is found not only between the L- and D-components (Figure \ref{fig6}) but also between the I- and D-components, whereas the I-component partly overlaps the D-component (Figure \ref{fig7}a).} In Figure \ref{fig6}, the L- and D-components show complementary spatial distribution. The L-component is distributed clearly along the edge of the D-component in the northeast and southwest of N44. The I- and D-components show {some} complementary spatial distribution, while the I-component partly overlaps with the D-component in the south of N44 D as shown in Figure \ref{fig7}b.

A possible interpretation of these spatial distributions is deceleration of the H{\sc i} gas by momentum conservation {in the interaction.} In the R136 region, we interpreted that the L-component penetrated into the D-component {with small deceleration when the D-component} has lower column density than the L-component, whereas the L- and D-components merge {together with significant deceleration} where they have {nearly the same} column density. The difference in the column density of colliding H{\sc i} flows then affects the velocity in the merged components by momentum conservation (Paper I). If we assume a similar collisional process, the distributions of H{\sc i} gas can be interpreted as follows; 
In the northeast and southwest of N44, {two parts of} the L-component are located where the intensity of the D-component is weaker than the surroundings (Figure \ref{fig6}). The average column density of the D-component in this region is $\sim$1.7$\times$10$^{21}$ cm$^{-2}$. It is likely that the L-component exhibits weakly a sign of deceleration, because $N$(H{\sc i}) of the L-component dominates that of the D-component when momentum conservation is considered. The initial column density of the L-component was probably high enough toward the region. 
In the south of N44, the I-component overlaps with the D-component which has the highest intensity of H{\sc i} (Figure \ref{fig7}a). The maximum column density of the D-component is $\sim$4.0$\times$10$^{21}$ cm$^{-2}$. We suggest that $N$(H{\sc i}) of the L- and D-components was nearly the same, and that the collision resulted in merging of the two components at an intermediate velocity of these components. The total $N$(H{\sc i}) of the merged component was then elevated to $\sim$6$\times$10$^{21}$ cm$^{-2}$ {due to the merging.} 
In the other regions, the I- and D-components show complementary spatial distribution, a characteristic of collision between the two components. The I-component exists but is weak in H{\sc i} in the north of N44B and C as shown in Figure \ref{fig7}a. This is probably because the stellar ionization already dispersed the colliding H{\sc i} gas. The age of N44 \citep[5--6 Myr;][]{1997A&AS..123..455W} is three times older than that of R136. We suggest the two components are colliding as well as in the other areas of N44.

Another possible interpretation is “displacement of the colliding two velocity components.” It is usually the case that the angle of the colliding two clouds {is not 0$\degr$ relative to the line of sight}. This collision angle $\theta$ makes a displacement of the complementary distribution between intensity depression {of the D-component} and the dense part of the I -component {according to the time elapsed since the initiation of the collision}. Thus, we are able to see the overlap of the I- and D-components in the south of N44 D shown in Figure \ref{fig7}a, {and the overlap is considered to be due to the projection}. In Figure \ref{fig7}b {where is applied a displacement of 107 pc as shown by an arrow in Figure \ref{fig7}.} We see the I-component better fits the edge of the D-component than in Figure \ref{fig7}a {and} they exhibit complementary distribution. 
In order to estimate the {optimum} displacement, {we applied the method which is presented in \cite{2018ApJ...859..166F}.} We calculated the overlap integral  $H$($\Delta$) in pc$^{2}$, which is a measure of the overlapping area of the strong I-component (Integrated intensity $>$ 550 K km s$^{-1}$) and the depression of the D-component (Integrated intensity$<$ 1100 K km s$^{-1}$). 
 
{We then searched for the best optimal solution of displacement to maximize the overlapping integral $H$($\Delta$) sweeping the I-component in position as a cross correlation function. The sweeping was done toward 5 different directions of $\Delta$-axis with the position angle from 170$\degr$ to 190$\degr$ with 5$\degr$ steps, and by the $\Delta$ offsets from $-400$ to $+400$ pc with 9.7 pc steps. It was then found that the $\Delta$-axis of 180$\degr$ and $\Delta$=107 pc gives the largest $H(\Delta)$ value as shown in Figure \ref{fig7}c. This displacement is indicated in Figure \ref{fig7}b by the arrow, and by the good complementary distributions between the strong I-component and the depressed D-component. More details about $H$($\Delta$) are given in Appendix of \cite{2018ApJ...859..166F}.}

The I-component partly overlaps with the D-component, for example at (R.A., Dec.)$\sim$ (5$^{h}$25$^{m}$0$^{s}$, $-$68$^{d}$20$^{m}$0$^{s}$) after {applying the displacement}. We note that the complementary distribution may not hold so strictly, depending on the three-dimensional spatial distributions of the D-component, and that the I-component may overlap in part the D-component, depending on the initial distribution of the two components.

These characteristics (Sections \ref{sec:1}--\ref{sec:3}) are consistent with the observational signatures of collision between the H{\sc i} flows shown in Paper I and {lend support} that the colliding H{\sc i} flows triggered the formation of $\sim$40 high-mass stars in N44 as in the R136 region.

\subsection{Stellar properties and {crossing} timescale}

In the previous section, we presented a scenario for massive star formation by colliding H{\sc i} flows in N44. In this sub-section, we discuss the stellar properties, the star formation rate, and the {crossing} timescale in the context of the scenario. {Crossing timescale is the time it takes for the L- or I-components to pass through the D-component in collision. }

\subsubsection{Distribution of high-mass stars and physical properties of H{\sc i} gas}\label{sec:4.2.1}
In Figure \ref{fig8}, we see that most of the high-mass stars are formed in the central region of N44 where H{\sc i} flow is colliding (Line B in Figure \ref{fig8}). We suggest that the high-mass stars were {mostly} formed by colliding H{\sc i} flows in Line B, if H{\sc i} flows are decelerated by the collision and {become} further compressed. In the Dec.-Velocity diagrams of the central region of N44 (Figure \ref{fig4}), the velocity separation between the two components in Line B is smaller than in {Lines A and C where the velocity separation is $\sim$50 km s$^{-1}$}, while in Line B the velocity separation is $\sim$20 km s$^{-1}$ . This difference in the velocity separation suggests that the L-component is {significantly} decelerated by the collision between the D-component in Line B, {leading to the most significant H{\sc i} gas compression.}

We calculated the total H{\sc i} mass and peak column density in Lines A, B, and C in Figure \ref{fig8} and did not find significant difference among the regions, {which suggests the current total $N$(H{\sc i}) is similar toward the three lines.} The age of N44 5--6 Myr \citep{1997A&AS..123..455W} makes it difficult to see the initial conditions {prior to the collision} because the H{\sc i} gas was significantly {dissipated} by the high mass stars in 5--6 Myrs. 
H{\sc i} gas within 25--30 pc of high-mass stars is dispersed by ionization and stellar winds at an empirical velocity of the ionization front of $\sim$5 km s$^{-1}$ observed in Galactic high mass clusters \citep{2016ApJ...820...26F}, and the amount of the H{\sc i} gas in Line B {may be less than the initial value}. Thus, H{\sc i} gas was possibly more concentrated on Line B than {Lines A and C before the collision}. A scenario that H{\sc i} gas was further compressed and high-mass stars formed in Line B is consistent with the observations.

We also estimated the total {Interstellar medium} mass of the star forming region N44 as illustrated by a white dashed circle in Figure \ref{fig2}b. The region is determined so that it includes the H$\alpha$ bubble. The mass of H{\sc i} is estimated to be $\sim$5$\times$10$^{5}$ $M_{\sun}$ by assuming the uniform column density of $\sim$6.0$\times$10$^{21}$ cm$^{-2}$ (see in Table {3}) and the total mass of H$_{2}$ is $\sim$5$\times$10$^{5}$ $M_{\odot}$ by using equation (2). We calculated the approximate total stellar mass of N44 ($M$*) as $\sim$1.0$\times$10$^{5}$ $M_{\sun}$ assuming the initial mass function (IMF) as follows. IMF varies with mass range of the star \citep{2001MNRAS.322..231K} as follows; ; $\phi$($M$) $dM$ $\propto$ $M^{-2.3}$ $dM$: $M$$>$0.5 $M_{\odot}$, $\phi$($M$) $dM$ $\propto$ $M^{-1.3}$ $dM$: $M$$>$0.5 $M_{\odot}$ $>$0.08 $M_{\odot}$, $\phi$($M$) $dM$ $\propto$ $M^{-0.3}$ $dM$: $M$$>$0.08 $M_{\odot}$ $>$ 0.01 $M_{\odot}$. When the stellar mass is larger than 0.5 $M_{\odot}$, we adopt $\alpha$=2.37 as the slope of IMF which obtained from BV photometry \citep{1997A&AS..123..455W}, and we adopt $\alpha$=1.37 as the slope of IMF for the stellar mass less than 0.5 $M_{\odot}$ \citep{2001MNRAS.322..231K}. The star formation efficiency (SFE) is about 10\% for total interstellar mass ($M$(H{\sc i}+H$_{2}$)), where we define SFE as $M$*/($M$*+$M$(H{\sc i}+H$_{2}$)).


%

\subsubsection{{Crossing timescale of collision}}
We found {a spatial} displacement between the I- and D-components shown in Figure \ref{fig7}b. We obtained the projected displacement to be 107 pc at a position angle of 180$\degr$ and the velocity separation is 20 km s$^{-1}$ from Figure \ref{fig3}. More details are given in section \ref{sec:4.2.1}. The {crossing} timescale is estimated to be $\sim$150 pc / 28 km s$^{-1}$$ \sim$5 Myr by assuming an angle of the cloud relative motion to the line of sight to be $\theta$ = 45$\degr$. The {crossing} timescale is consistent with the age of N44 \citep{1997A&AS..123..455W}.

We also estimated the crossing timescale of the collision from a ratio of the cloud size and the relative velocity between the two clouds. The apparent spatial extent of the colliding clouds forming high-mass stars is about 200 to 300 pc in Figures \ref{fig5}c. The velocity separation is $\sim$28 km s$^{-1}$ if we assume the relative motion between the two clouds has an angle of 45$\degr$. We calculated a ratio of the cloud size to the relative velocity, giving $\sim$7 to 10 Myr as the crossing timescale. The time scale gives a constraint on the upper limit for the age of the high-mass stars formed by colliding H{\sc i} flows. From these results, we suggest that the {crossing} timescale is 5--10 Myr, {while a younger age is more likely because the deceleration tends to lengthen the timescale than the actual value.}

\subsection{Inflow of the metal poor gas from the SMC}
We found evidence of {H{\sc i} gas} inflow from the SMC from {the H{\sc i} and $Planck$/$IRAS$ data} (Paper I), and suggest that the H{\sc i} gas of the SMC is possibly mixed with the H{\sc i} gas of the LMC {by the tidal interaction.} Specifically, we derived a gas-dust ratio from a scatter plot between the dust optical depth at 353 GHz ($\tau_{353}$) and the 21 cm H{\sc i} intensity ($W$(H{\sc i})) in the LMC. We used the data with the highest dust temperature which are most probably optically thin H{\sc i} emission \citep{2014ApJ...796...59F,2015ApJ...798....6F}. The slope of the N44 region in the plot is steeper by $\sim$30\% than that of the {Bar} region which is {an older stellar system than the H{\sc i} Ridge.} This difference is interpreted as due to different dust abundance between them.

The results of numerical simulations \citep{2007MNRAS.381L..16B} show that the H{\sc i} gas of the SMC was mixed with the LMC gas by the close encounter happened {0.2 Gyr ago}. If a mixture of the SMC gas and the LMC gas was triggered by the tidal interaction, the N44 region should contain low-metallicity H{\sc i} gas of the SMC. If the H{\sc i} gas of the N44 region consists of H{\sc i} mass both of the SMC and the LMC at a ratio of 3:7 by assuming that sub-solar metallicity of 1/10 $Z_{\odot}$ in the SMC \citep{2003A&A...400...21R} and of 1/2 $Z_{\odot}$ in the LMC \citep{2002A&A...396...53R}, the different slopes in Figure \ref{fig11} is explained by the mixing. This scenario deserves further {pursuit} by {more elaborated} numerical simulations of the hydrodynamical tidal interaction into detail.




\subsection{Comparison with the H{\sc i} Ridge and {H{\sc i} Ridge including} R136}
We compared physical parameters of N44 and {H{\sc i} Ridge} as summarized in Table {3}. We see the different physical conditions and the activity of high-mass star formation between N44 and {H{\sc i} Ridge}. {H{\sc i} Ridge contain massive stars of R136 in the core of 30 Doradus.} The number of high-mass stars of N44 is 10\% of {H{\sc i} Ridge \citep{2013A&A...558A.134D}} and the H{\sc i} intensity of the L-component toward N44 is 25\% of the H{\sc i} Ridge. The spatial extent of the colliding H{\sc i} flow of N44 is $\sim$25\% of the H{\sc i} Ridge. In addition, we find the I-component toward N44, which has smaller velocity separation than the L-component. 

Furthermore, the timescale of the collision is different between N44 and R136. The {crossing} timescale of N44 is 5--6 Myr, but that of R136 is $\sim$3 Myr (derived from Paper I). This difference may have been caused by {difference in} timing of the inflow {impact of the tidally driven} SMC gas \citep{2007MNRAS.381L..16B}.

From these results, we speculate that some physical parameters of the collision, for instance, the amount of the inflow gas from the SMC and the velocity of collision determine the activity of high-mass star formation. However, it is not understood which parameter is important, because we do not have a large number of observational examples {and appropriate numerical simulations of the colliding H{\sc i} flows. We need to study other star forming regions in the LMC as shown in Figure \ref{fig1} and to carry out statistical analyses more extensively as a future work.}

\section{Conclusion}\label{sec:con}
The main conclusions of the present paper are summarized as follows; 
\begin{enumerate}
\item We analyzed the high spatial resolution H{\sc i} data (1 arcmin; corresponding to $\sim$15 pc at a distance of the LMC) observed by the ATCA \& Parkes telescope \citep{2003ApJS..148..473K}, and revealed the spatial and velocity distributions of N44. We {confirmed} that the two H{\sc i} components at different velocities, the L- and D-components  \citep{2017PASJ...69L...5F}, are colliding at a 500 pc scale with velocity difference of $\sim$30 to 60 km s$^{-1}$. The collision is characterized by the spatial complementary distribution and bridge features in velocity space. {In addition, we newly defined the I (intermediate)-component between the two velocities as a decelerated gas component due to the collision.}

\item We calculated total mass of massive stars ({WR and O-type stars}) and total mass of the H{\sc i} gas, and derived star formation efficiency in N44 to be $\sim$10\% from a ratio of [the stellar mass]/[the stellar mass plus the H{\sc i} and CO mass]. A timescale of the collision is estimated by the displacement of the I-component and the velocity difference. We estimated that $\sim$5 Myr passed after the collision by assuming that the I-component moved by $\sim$100 pc in projection at a velocity of 20 km s$^{-1}$. This timescale is consistent with an age of the star forming region in N44 \citep{1997A&AS..123..455W}. These results suggest it possible that the high-mass stars in N44 were formed by triggering in the colliding H{\sc i} flows.

\item We found that the L-component is metal poor as indicated by the low dust optical depth. We estimated the metal content by a comparison of dust optical depth at 353 GHz ($\tau_{353}$) measured by the $Planck$/$IRAS$ and the intensity of H{\sc i} ($W$(H{\sc i})). We derived an indicator of gas-dust ratio ($W$(H{\sc i}) /$\tau_{353}$) by the least-squares {fitting} which is assumed to have a zero intercept. The gas-dust ratio ($W$(H{\sc i}) /$\tau_{353}$) of N44 is by 30\% larger than that of the Bar region. If the N44 region consists of H{\sc i} mass from the SMC \citep[1/10 $Z_{\odot}$;][]{2003A&A...400...21R} and LMC \citep[1/2 $Z_{\odot}$;][]{2002A&A...396...53R} at a ratio of 3:7, slopes different by 30\% in Figure \ref{fig11} are explained. This result supports the tidal origin of the H{\sc i} flow and mixing of the SMC gas with the LMC gas as theoretically predicted \citep{2007MNRAS.381L..16B}. The H{\sc i} Ridge is mixed by the H{\sc i} gas of the SMC and LMC at a ratio of 1:1 (Paper I) and it means the influence of the tidal interaction was weaker in N44 than in the H{\sc i} Ridge.  {Since the larger dispersion of the scatter plot between $\tau$$_{353}$ and $W$(H{\sc i}) in the Bar region could be results of metal enrichment by star formation history. More detail study of gas-to-dust ratio for the entire LMC will be presented in the upcoming paper (K. Tsuge et al., in preparation). }

\item We compared a collision size scale of the N44 region and that of the H{\sc i} Ridge, and found that the collision of the N44 region is of a smaller-scale than that of the H{\sc i} Ridge. Specifically, a spatial extent of the colliding H{\sc i} flow is about 1/4 of the H{\sc i} Ridge and the intensity of the L-component is weaker by 1/4 than that of the H{\sc i} Ridge. It is clearly seen in N44 that the spatial scale of collision, the intensity of blue-shifted H{\sc i} gas, the number of the formed massive stars, and the H{\sc i} mass of the inflow from the SMC are smaller than in the H{\sc i} Ridge.

\item We discussed a scenario that a tidally-driven colliding H{\sc i} flow triggered the formation of {high-mass} stars based on the present results. The blue-shifted H{\sc i} components (the L- and I -velocity components) were formed by a tidal stripping between the LMC and SMC about 0.2 Gyrs ago \citep{2007PASA...24...21B}. The perturbed H{\sc i} gas is colliding with the H{\sc i} gas in the LMC disk at present. This collision formed $\sim$40 O/WR stars in N44 by the same mechanism as in the R136 region (Paper I), but the number of high mass stars of N44 is $\sim$1/10 as compared with that in R136. The spatial scale of collision and the H{\sc i} intensity of the blue-shifted components are smaller than H{\sc i} Ridge, probably because the tidal interaction between the LMC and SMC was weak in the N44 region. More details of the collision are yet to be better clarified for understanding the cause of the difference, and we need to increase the number of the cases where the colliding H{\sc i} flows trigger high mass star formation in the LMC.


\end{enumerate}

\acknowledgments
The NANTEN project is based on a mutual agreement between Nagoya University and the Carnegie Institution of Washington (CIW). We greatly appreciate the hospitality of all the staff members of the Las Campanas Observatory of CIW. We are thankful to many Japanese public donors and companies who contributed to the realization of the project. This study was financially supported by JSPS KAKENHI Grant Number 15H05694. This work was also financially supported by Career Development Project for Researchers of Allied Universities. The ATCA, Parkes, and Mopra radio telescope are part of the ATNF which is funded by the Australian Government for operation as a National Facility managed by CSIRO. The UNSW Digital Filter Bank used for the observations with the Mopra Telescope was provided with support from the Australian Research Council. Based on observations obtained with Planck, an ESA science mission with instruments and contributions directly funded by ESA Member States, NASA, and Canada. The Southern H-Alpha Sky Survey Atlas, which is supported by the National Science Foundation.
Cerro Tololo Inter-American Observatory (CTIO) is operated by the Association of Universities for Research in Astronomy Inc. (AURA), under a cooperative agreement with the National Science Foundation (NSF) as part of the National Optical Astronomy Observatories (NOAO).  The MCELS is funded through the support of the Dean B. McLaughlin fund at the University of Michigan and through NSF grant 9540747.

\clearpage
\appendix
\section{Decompositions of the L- and D-components of H{\sc i}}
To decompose the L- and D-components of H{\sc i} in the LMC, we used the following method.

\begin{enumerate}

\item {Verification of spectrum} 
We investigate the velocity structures of H{\sc i} spectra of the whole LMC by eye. We found that H{\sc i} spectra {are dominated by single-peaked component, with multiple component distributed in some regions.}

\item {Spectral fitting}

\begin{enumerate}
\renewcommand{\labelenumii}{\arabic{enumii}).}
\item Searching for a velocity channel that has the strongest intensity of H{\sc i} spectrum (Figure \ref{figA1}a).  We define emission peak with the brightness temperature higher than 20 K as the significant detection. 
\item Fitting the H{\sc i} profile to a Gaussian function. The velocity range used for the fitting is $\pm$10 channels ($\sim$16.5 km s$^{-1}$) with respect to the central velocity shown in Figure \ref{figA1}a.

\item The fitting was performed to single Gaussian distribution as follow; 
\begin{eqnarray}
f(x)=A_{0}  \ {\rm exp} \left(-\frac{(v-v_{0})^{2}}{2\sigma^{2}}\right),
\end{eqnarray}

where $A_{0}$, $v_{0}$, and $\sigma$ are the height, the central velocity,  and the velocity dispersion of the Gaussian, respectively set to be free parameters. 


\item Subtracting Gaussian distribution that obtained by 2) from the original H{\sc i} spectrum as the residual is shown in Figure \ref{figA1}b.

\item Repeat 1)--4) steps for each spectrum twice. 
\end{enumerate}

\item Comparison of the peak velocity

We compared central velocities of the first and second peaks we identified by the above steps for each pixel, and defined the L-component as the components having lower velocity and the D-component having higher velocity. If the above procedure only gives a single velocity component, it is attributed to the D-component. 
Then, we obtained the central velocity distributions of the L- and D-components.

The velocity distribution of the D-component indicates the rotation of the LMC (Figure \ref{figA2}a). However, we see some local velocity structures {differing from the galaxy rotation perturbed by }such as SNRs, super bubbles, and supergiant shells. So, we took mean velocity of 500 pc$\times$500 pc area by using the median filter for each pixel to remove these influence. Figure \ref{figA2}b shows the smoothed rotation velocity map of the LMC. The disk of the LMC rotate to the west from the east and this result consists with \cite{2003ApJS..148..473K}.

\item Subtraction of the galactic rotation velocity

To obtain the velocity relative to the disk rotation velocity ($V_{\rm offset}$) for each pixel, we subtracted the thus smoothed radial velocity of the D-component ($V_{\rm D}$) from the original velocity ($V_{\rm LSR}$) by shifting the velocity channels.

\begin{eqnarray}
V_{\rm offset}=V_{\rm LSR}-V_{\rm D}
\end{eqnarray}

\end{enumerate}
\clearpage

\section{Rotation curve of the D-component}

\begin{figure*}[htbp]
\begin{center}
\includegraphics[width=15cm]{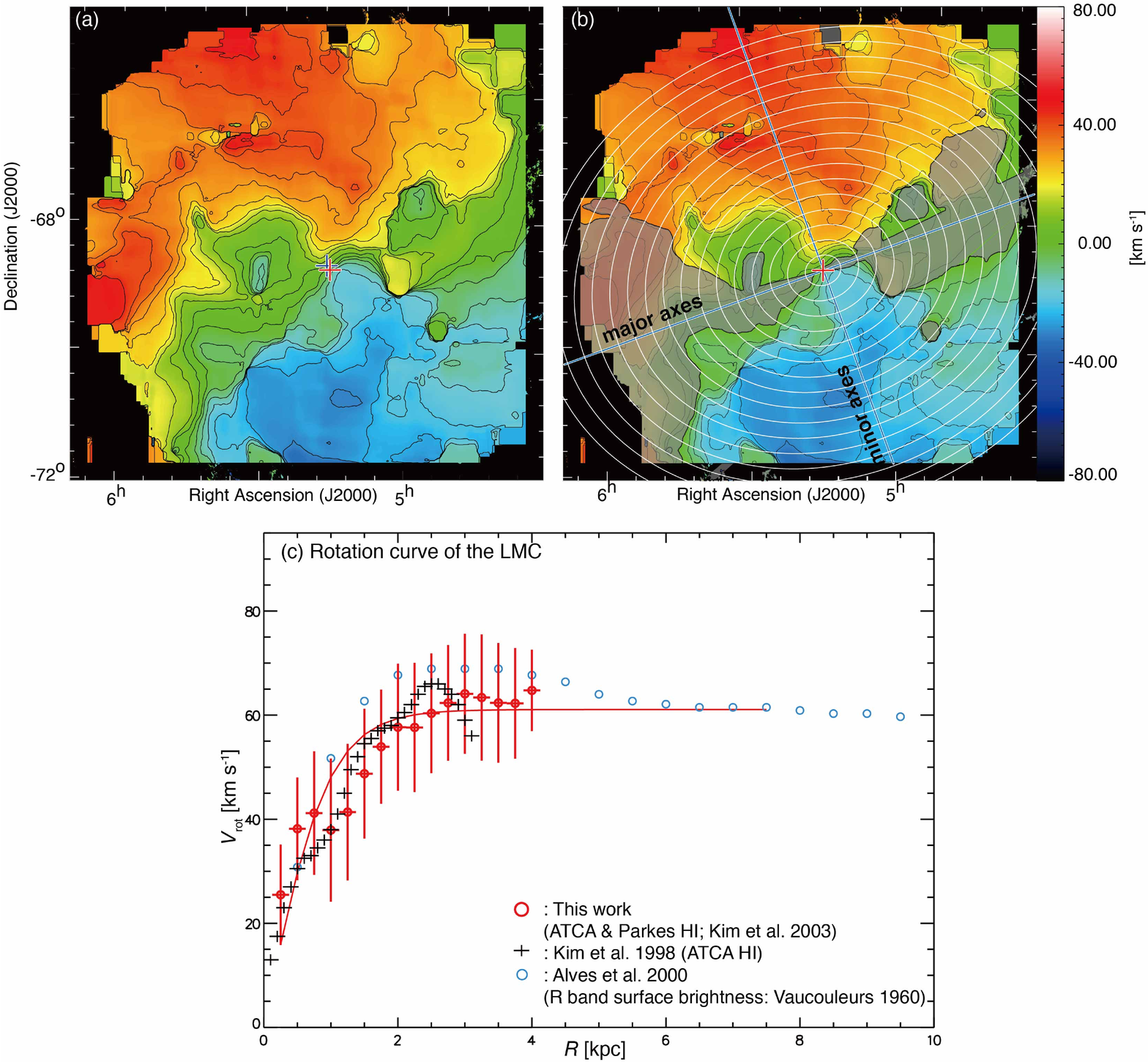}
\end{center}
\caption{(a) Radial velocity field of the D-component relative to the system velocity of the LMC. This velocity field is smoothed by the median filter with the size of filter of 500 pc $\times$ 500 pc. The lowest contour level and intervals are $-$50 km s$^{-1}$ and 5 km s$^{-1}$, respectively. The blue and red crosses indicate the positions of kinematic center derived from \cite{1998ApJ...503..729K} and this study, respectively. The lowest contour level and intervals are 0.25 kpc and 0.25 kpc, respectively. (b) Same image of (a) but superposed with the mask regions (shaded areas). Black contours and red cross are the same as in (a). White ellipses indicate the projected concentric circles around the kinematic center (red cross) on the plane of the galaxy. The cyan lines indicate major and minor axis of the ellipses. (c) Rotation curve of the LMC. The horizontal and vertical axis are radius on the plane of the galaxy and rotation velocity at the radius, respectively. Results of this study are plotted with red open circles. The black crosses indicate the H{\sc i} data of \cite{1998ApJ...503..729K}. The cyan open circle indicate stellar disk rotation curve derived from the $R$--band surface brightness data of \cite{2000ApJ...542..789A}. The mass--to--luminosity ratio is assumed to be constant of 2.2. }
\label{figB1}
\end{figure*}

\begin{figure*}[htbp]
\begin{center}
\includegraphics[width=\linewidth]{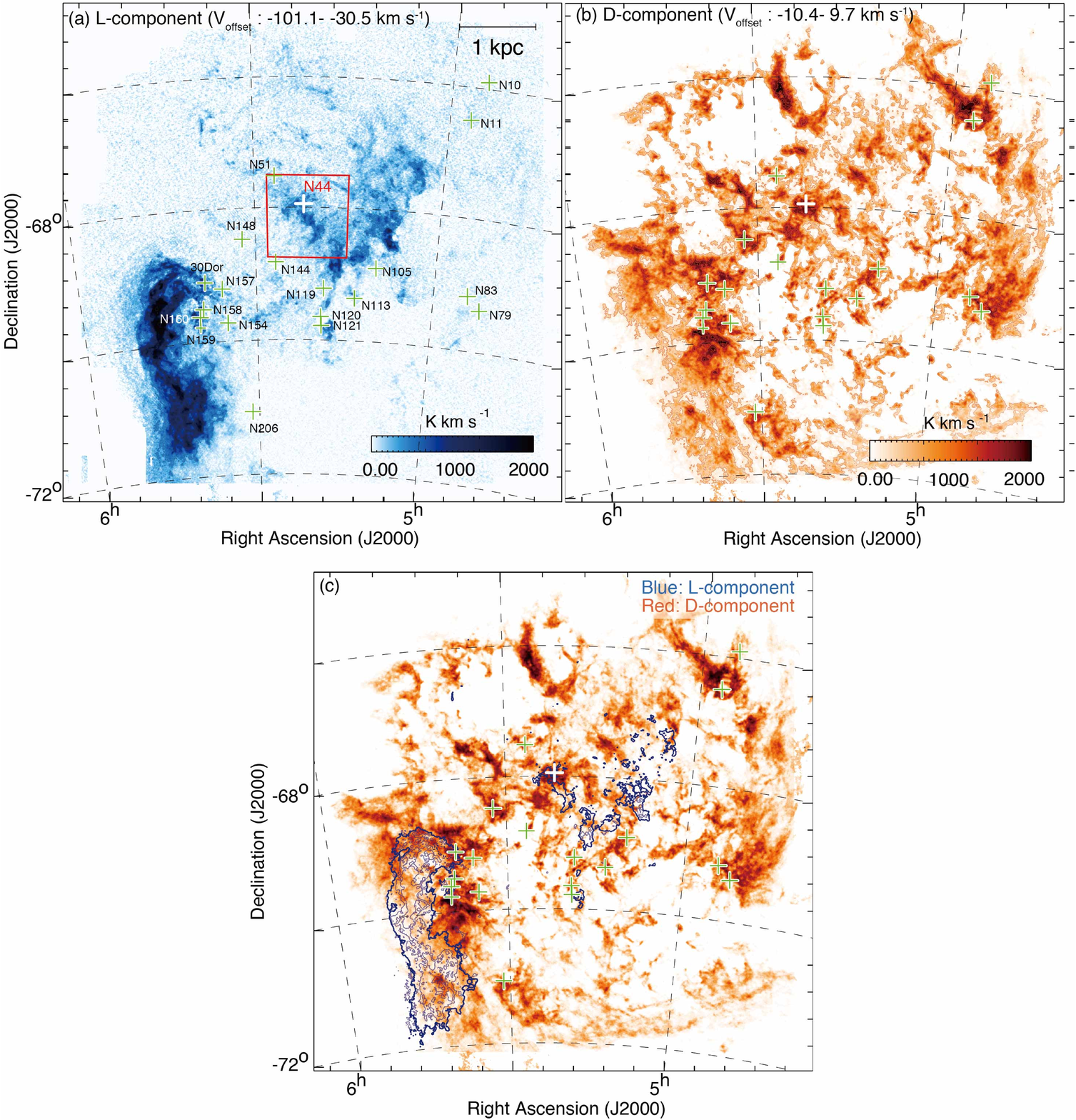}
\end{center}
\caption{The H{\sc i} integrated intensity maps of (a) the L-component and (b) the D-component. The integration velocity range is $V_{\rm offset}$= $-$101.1 -- 30.5 km s$^{-1}$ for the L-component and is $V_{\rm offset}$ =$-$10.4 --9.7 km s$^{-1}$  for the D-component. Crosses show the positions of luminous H{\sc ii} regions \citep[$S_{H\alpha}$  $>$ 1$\times$10$^{-4}$ erg cm$^{-2}$ s$^{-1}$ sr$^{-1}$][]{2016MNRAS.457.2048A}. The red box in (a) denotes the N44 H{\sc ii} region. The contour levels are 350, 800, 1200, and 2400 K km s$^{-1}$  for (a); 400, 800, 1200, 1600, 2000, and 2400 K km s$^{-1}$ for (b)). 
(c) H{\sc i} intensity map of the L-component by contours superposed on the D-component image. The velocity range and contour levels are the same as in (a) and (b).}
\label{fig1}
\end{figure*}

\begin{figure*}[htbp]
\begin{center}
\includegraphics[width=\linewidth]{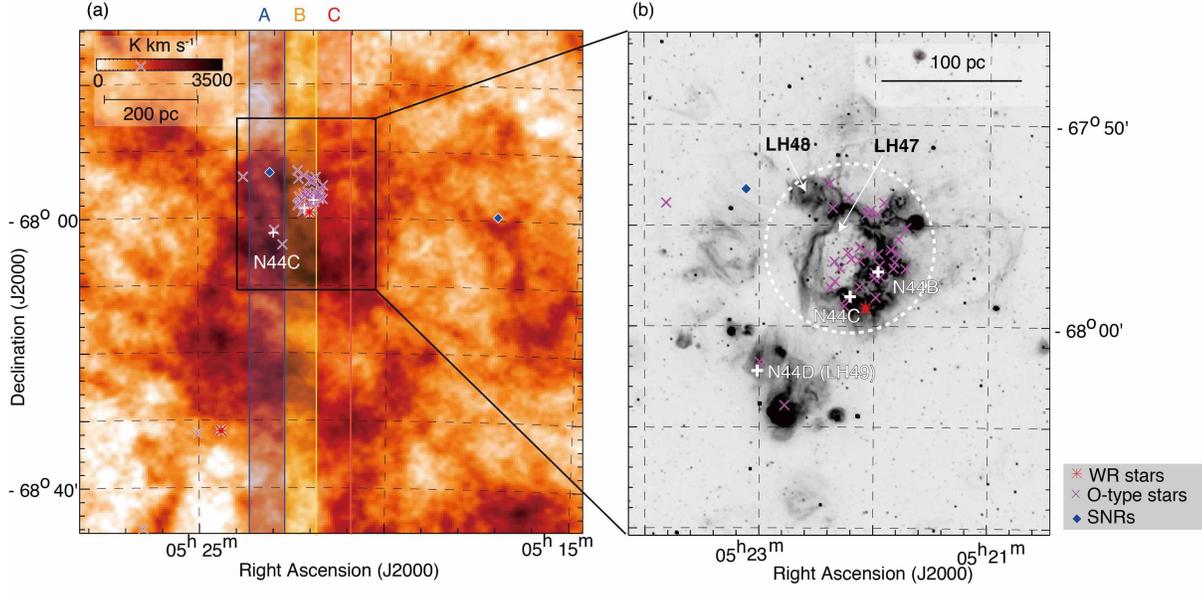}
\end{center}
\caption{(a)H{\sc i} total intensity map toward N44. The integration velocity range is $V_{\rm offset}$ = $-$109.8--89.7 km s$^{-1}$ . Positions-velocity diagrams along the Line A, B, and C passing through the east, center and west of the N44 region are demonstrated in Figure \ref{fig4}, respectively. (b) An enlarged H$\alpha$ image of N44 region. Image is H$\alpha$ \citep[MCEL;][]{1999IAUS..190...28S}. The red asterisks, pink crosses, and blue diamonds indicate WR stars, O-type stars {\citep{2009AJ....138.1003B}}, and SNRs \citep{2016A&A...585A.162M}, respectively.}
\label{fig2}
\end{figure*}

\begin{figure*}[htbp]
\begin{center}
\includegraphics[width=\linewidth]{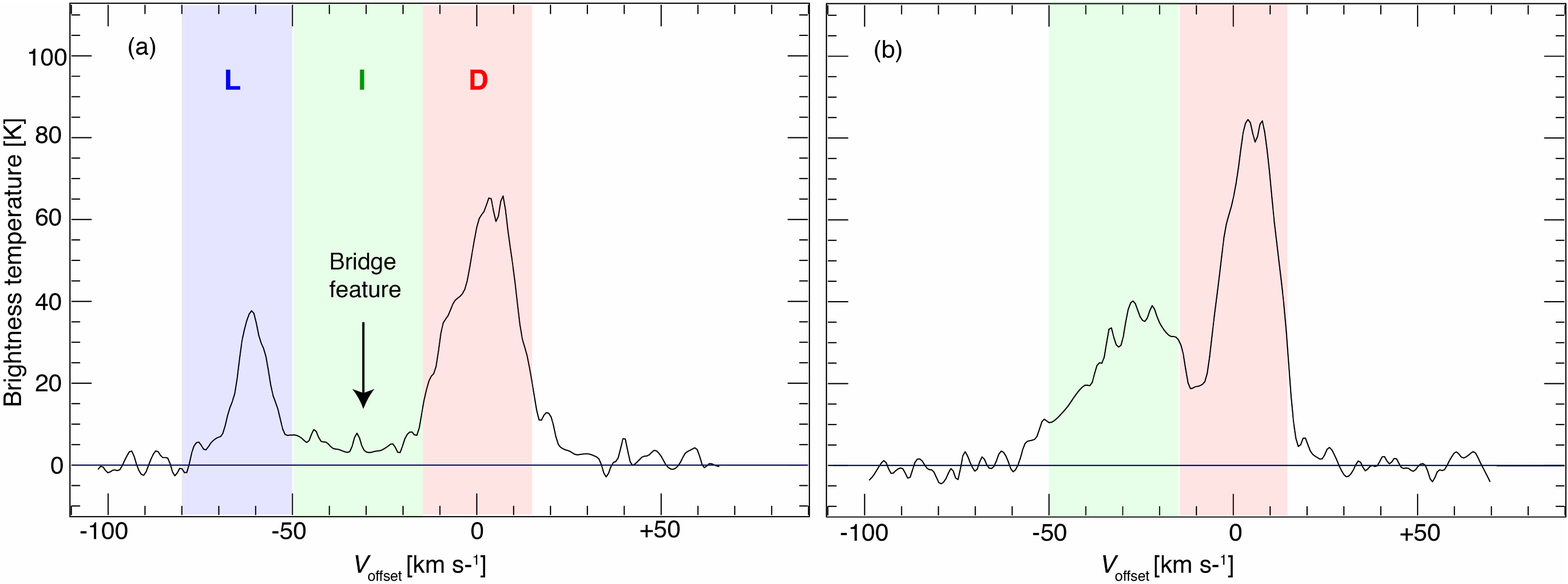}
\end{center}
\caption{The typical spectra of H{\sc i} at (a) (R.A., Dec.)=(5$^{h}$22$^{m}$59.23$^{s}$, $-$67$^{d}$55$^{m}$2.12$^{s}$) and (b) (R.A., Dec.)=(5$^{h}$22$^{m}$23.20$^{s}$, $-$68$^{d}$6$^{m}$9.02$^{s}$). The velocity ranges of shaded area correspond to the integration ranges as shown in Figure \ref{fig5}.}
\label{fig3}
\end{figure*}

\begin{figure*}[htbp]
\begin{center}
\includegraphics[width=\linewidth]{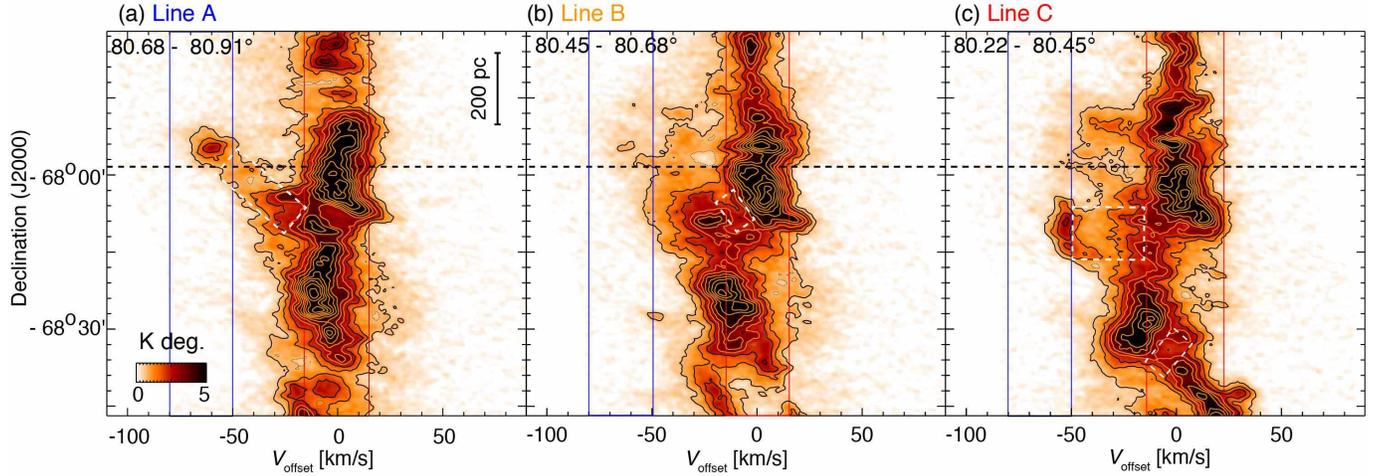}
\end{center}
\caption{The H{\sc i} Dec.-velocity diagrams along the lines A, B, and C. The integration is in R.A. over the range from 80$\fdg$68 to 80$\fdg$91 for (a); 80$\fdg$45 to 80$\fdg$68 for (b); 80$\fdg$22 to 80$\fdg$45 for (c). The lowest contour level and intervals are 0.7 K deg. and 0.7 K deg., respectively. Black dashed lines indicate the position of N44C. The white boxes in (a), (b), and (c) indicate the positions of bridge features.}
\label{fig4}
\vspace*{-0.5cm}
\end{figure*}

\begin{figure*}[htbp]
\begin{center}
\includegraphics[width=\linewidth]{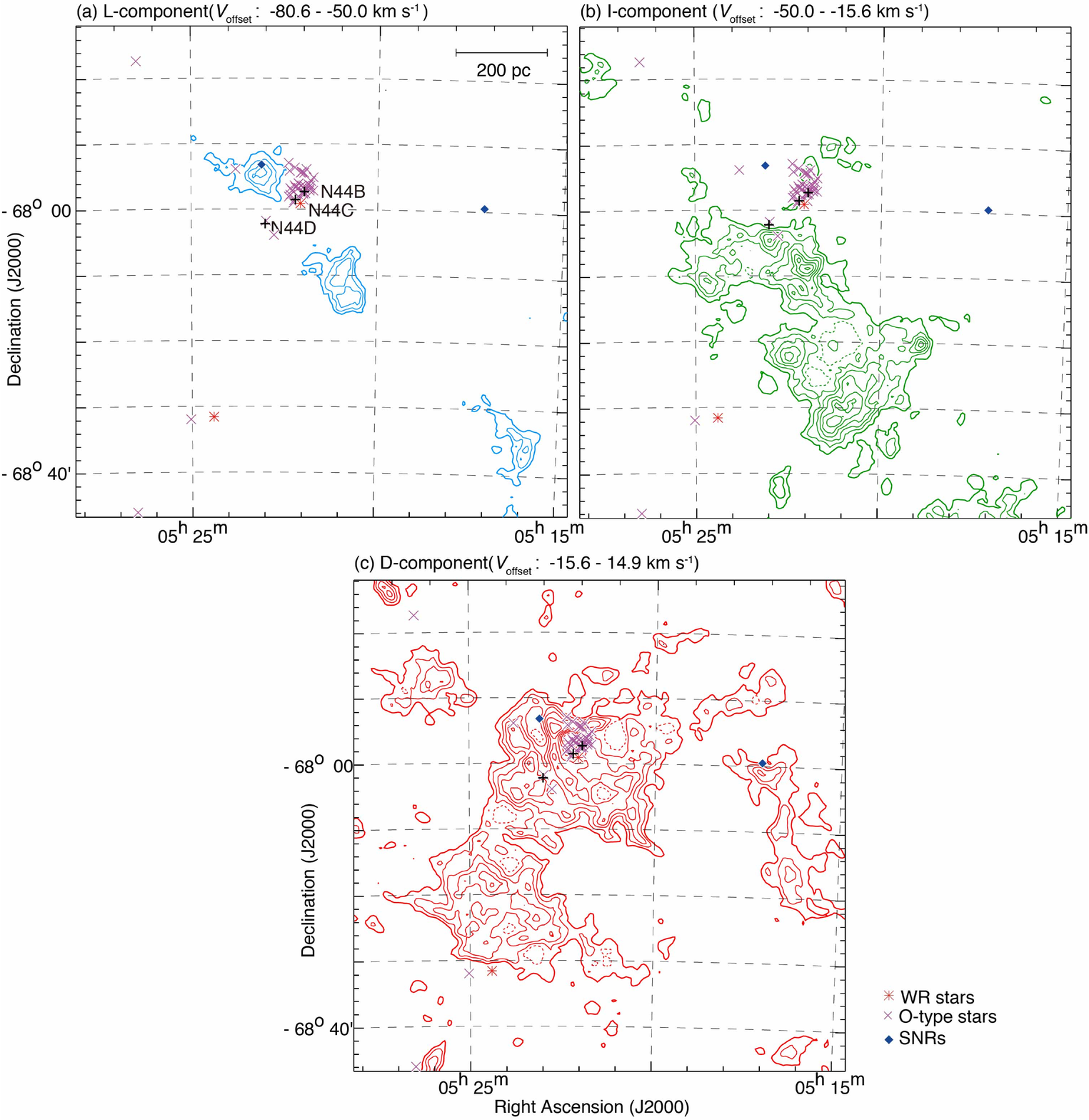}
\end{center}
\caption{Integrated intensity distributions of H{\sc i} for the three velocity components shown in Figure 3. The integration range is $V_{\rm offset}$: $-$80.6-- $-$50.0 km s$^{-1}$ for (a); $V_{\rm offset}$: $-$50.0-- $-$15.6 km s$^{-1}$ for (b); $V_{\rm offset}$: $-$15.6--9.7 km s$^{-1}$ for (c). The lowest contour level and intervals are 200 K km s$^{-1}$ and 100 K km s$^{-1}$ for (a); 500 K km s$^{-1}$ and 150 K km s$^{-1}$ for (b); 1000 K km s$^{-1}$ and 200 K km s$^{-1}$ for (c). The symbols are the same as in Figure \ref{fig2} {(c).}}
\label{fig5}
\end{figure*}

\begin{figure*}[htbp]
\begin{center}
\includegraphics[width=12cm]{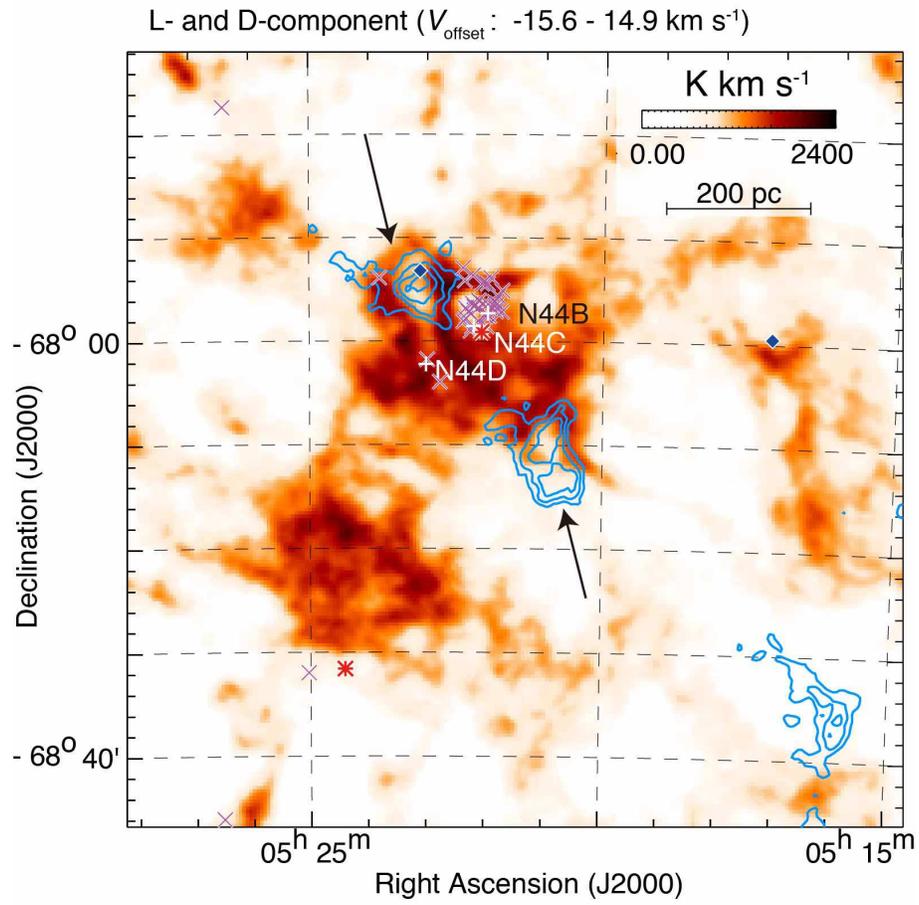}
\end{center}
\caption{H{\sc i} intensity maps of the L-component by contours superposed on the D-component by image. The contour levels and symbols are the same as in Figure \ref{fig2}.}
\label{fig6}
\end{figure*}

\begin{figure*}[htbp]
\begin{center}
\includegraphics[width=15cm]{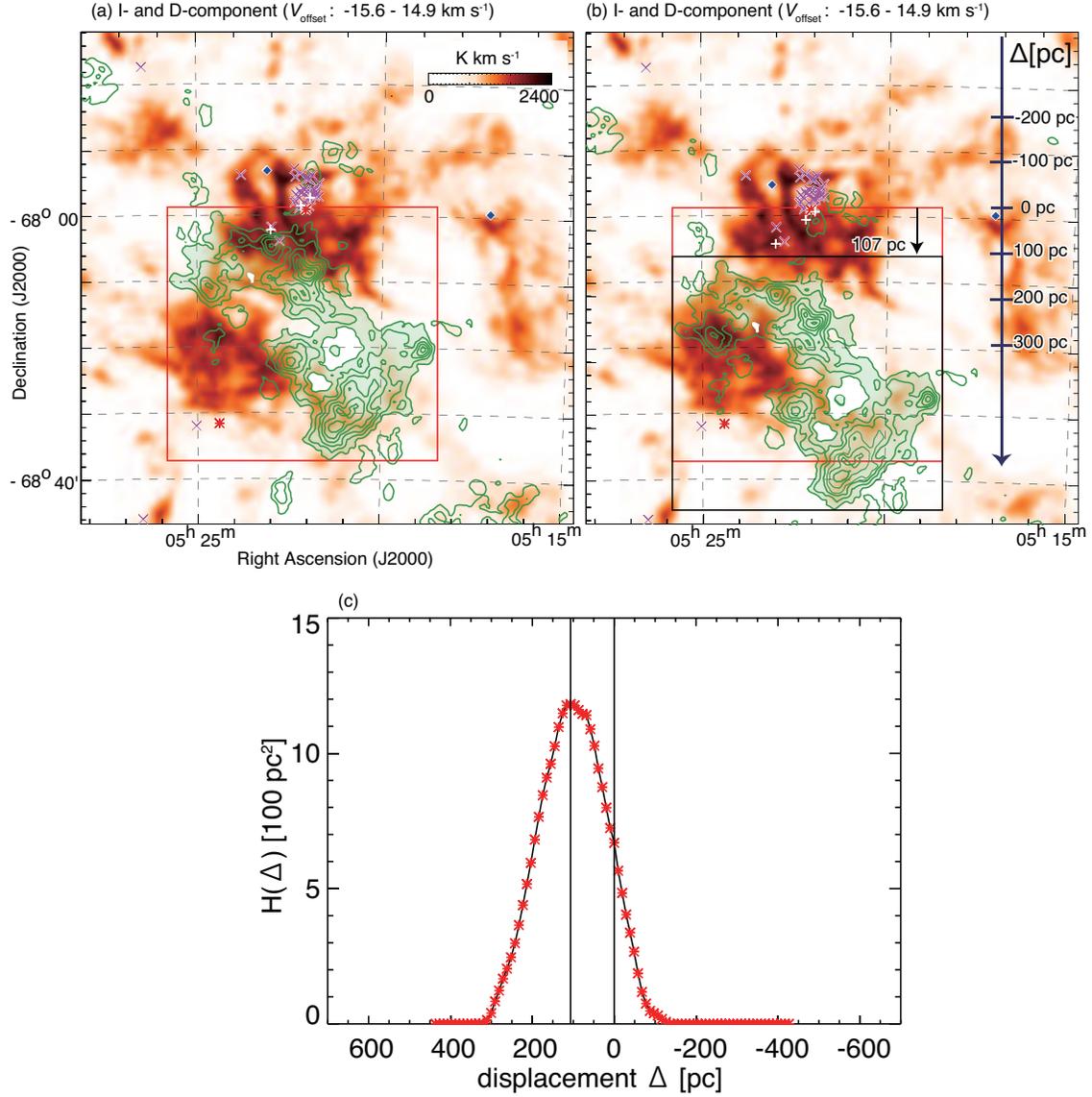}
\end{center}
\caption{(a) H{\sc i} intensity maps of the I-component by contours superposed on the D-component by image. (b) Same as (a), but the contours of the I-component are displaced. The projected displacement is 107 pc with a position angle of 180 deg. The red box denotes the initial position of the I-component, and the black box of (b) denotes the displaced position. The green shaded regions of (a) and (b) illustrate the H{\sc i} intensity of 500 K km s$^{-1}$ or higher. The contour levels and symbols are the same as in Figure \ref{fig2}. {The $\Delta$-axis = 180 deg was taken as the optimum value for the direction of the displacement of the I-component. The origin of $\Delta$-axis denotes the initial position of the I-component. (c) The overlapping integral $H(\Delta)$ in pc$^{2}$ defined in Sec 4.1.3. The strong I-component ($>$ 550 K km s$^{-1}$) and the depressed D-component ($<$ 1100 K km s$^{-1}$) were assumed to be a uniform value of 1.0 (arbitrary unit). The horizontal axis is the value displacement $\Delta$ [pc] shown in (b).}}
\label{fig7}
\end{figure*}

\begin{figure*}[htbp]
\begin{center}
\includegraphics[width=11cm]{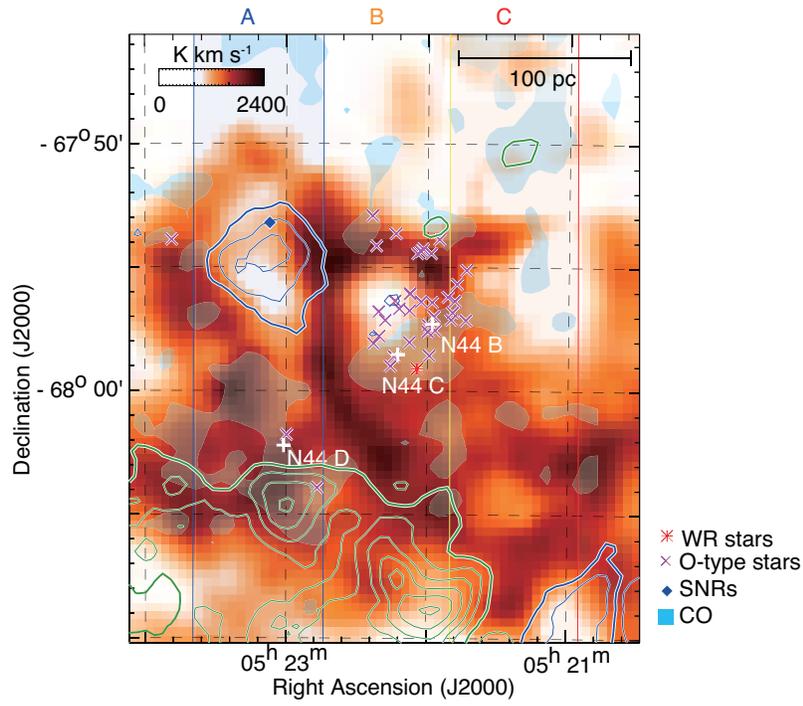}
\end{center}
\caption{The intensity map of H{\sc i} consisting of three velocity components around N44. The image indicates the D-component ($-$15.6--14.9 km s$^{-1}$). The blue and green contours indicate L- and I-components, respectively. The contour levels and symbols are the same as in Figure \ref{fig5}. The light-blue shaded area indicate the integrated intensity map of CO \citep[$>$ 3$\sigma$ (3.5 K km s$^{-1}$);][]{2011ApJS..197...16W} with the integration range of $V_{\rm offset}$: $-$109.8--89.7 km s$^{-1}$}
\label{fig8}
\vspace*{-0.5cm}
\end{figure*}

\begin{figure*}[htbp]
\begin{center}
\includegraphics[width=\linewidth]{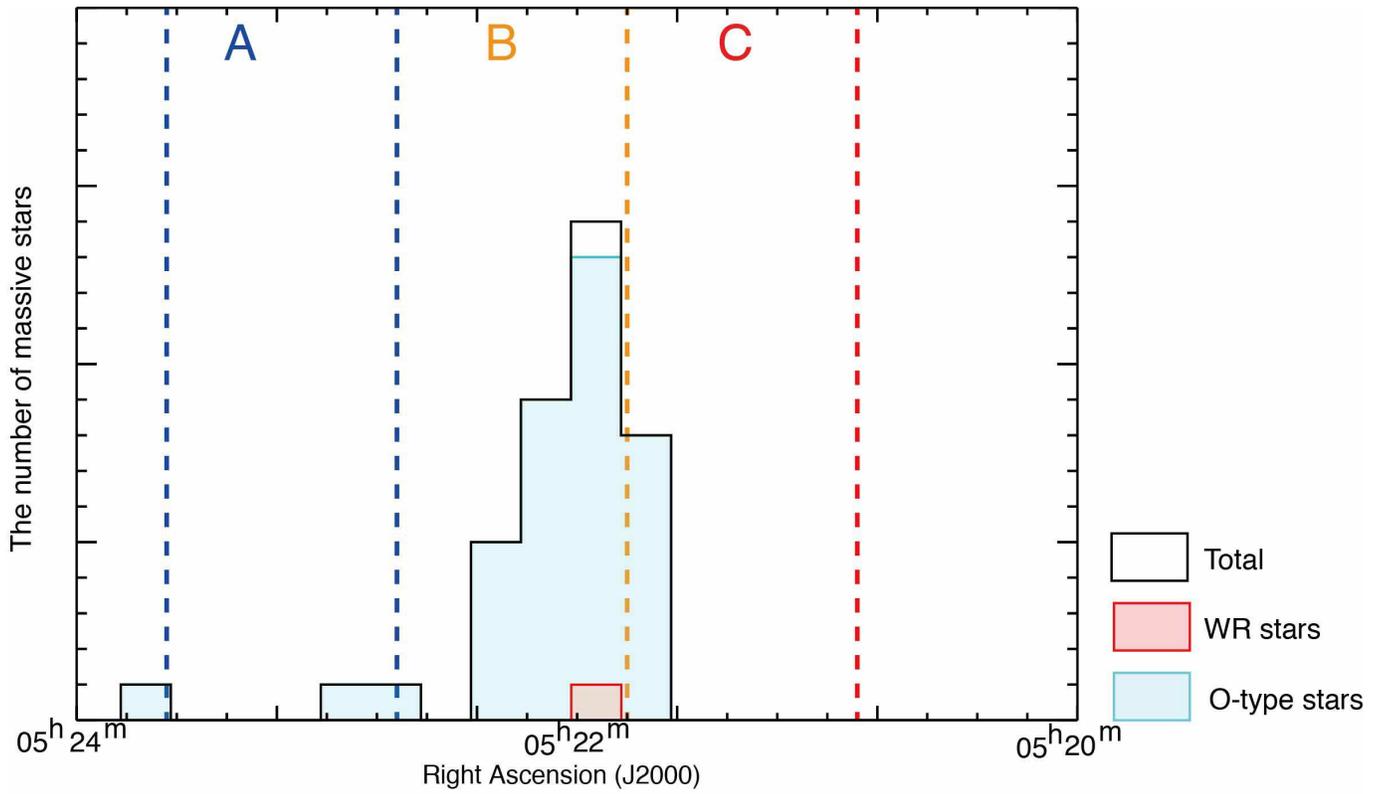}
\end{center}
\caption{Histgrams of the number of massive stars. The horizontal and vertical axis are R.A. and the number of massive stars located within the each R.A. bins, respectively. The histograms of the number of WR stars, O-type stars, and the total number of massive stars are shown by red- and blue-shaded areas, and open area, respectively.}
\label{fig9}
\vspace*{-0.5cm}
\end{figure*}

\begin{figure*}[htbp]
\begin{center}
\includegraphics[width=\linewidth]{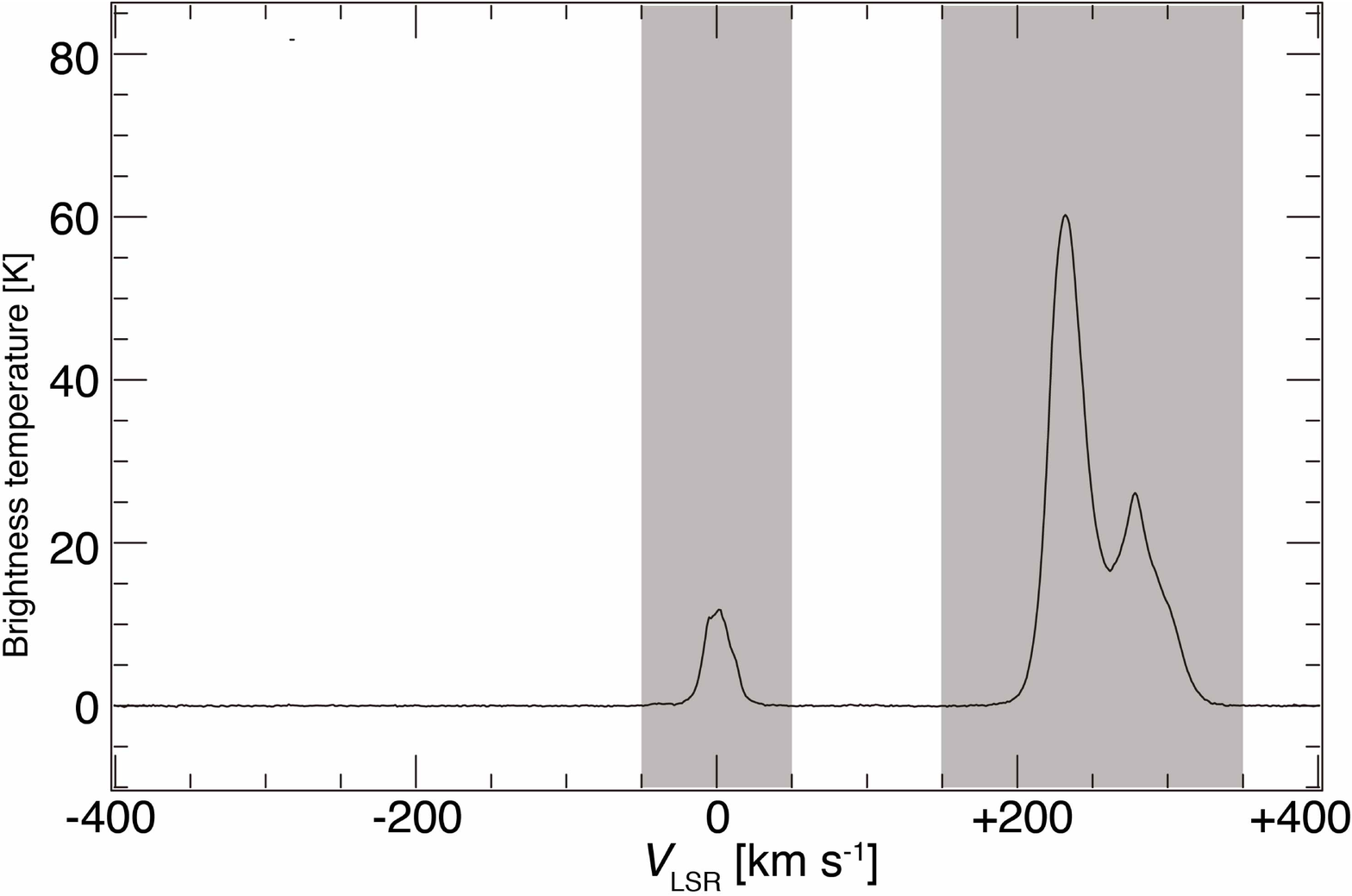}
\end{center}
\caption{The typical profile of H{\sc i} spectrum \citep{2016A&A...594A.116H} at (R.A., Dec.)=(5$^{h}$45$^{m}$39.09$^{s}$, $-$69$^{d}$35$^{m}$41.68$^{s}$).}
\label{fig10}
\end{figure*}

\begin{figure*}[htbp]
\begin{center}
\includegraphics[width=\linewidth]{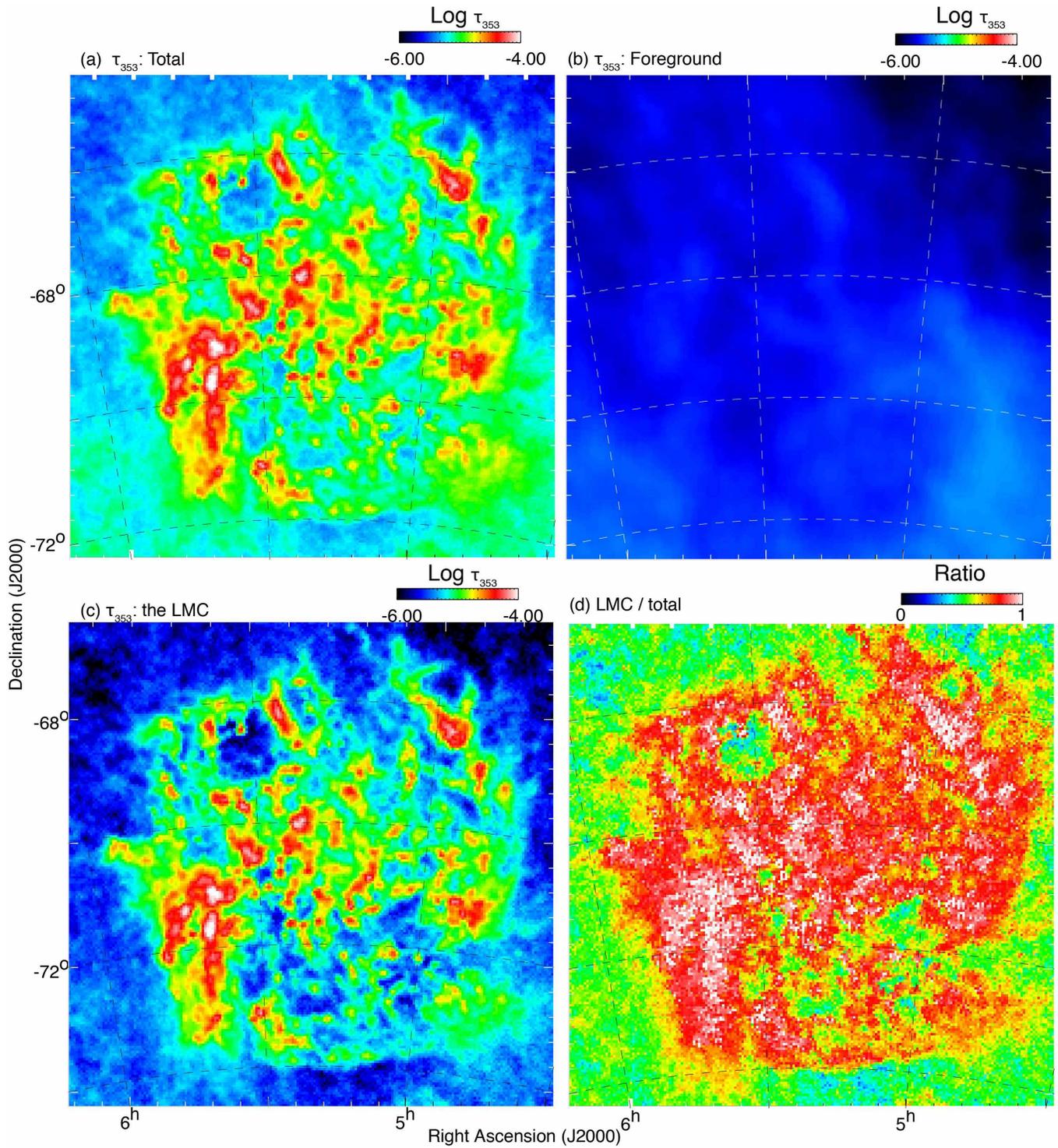}
\end{center}
\caption{Spatial distribution of $\tau$$_{353}$ toward the LMC. (a), (b), (c), and (d) indicate total quantity of $\tau$$_{353}$, the Galactic foreground component of $\tau$$_{353}$, the LMC component of $\tau$$_{353}$, and the ratio of the LMC component to the total $\tau$$_{353}$, respectively.}
\label{tau}
\end{figure*}

\begin{figure*}[htbp]
\begin{center}
\includegraphics[width=\linewidth]{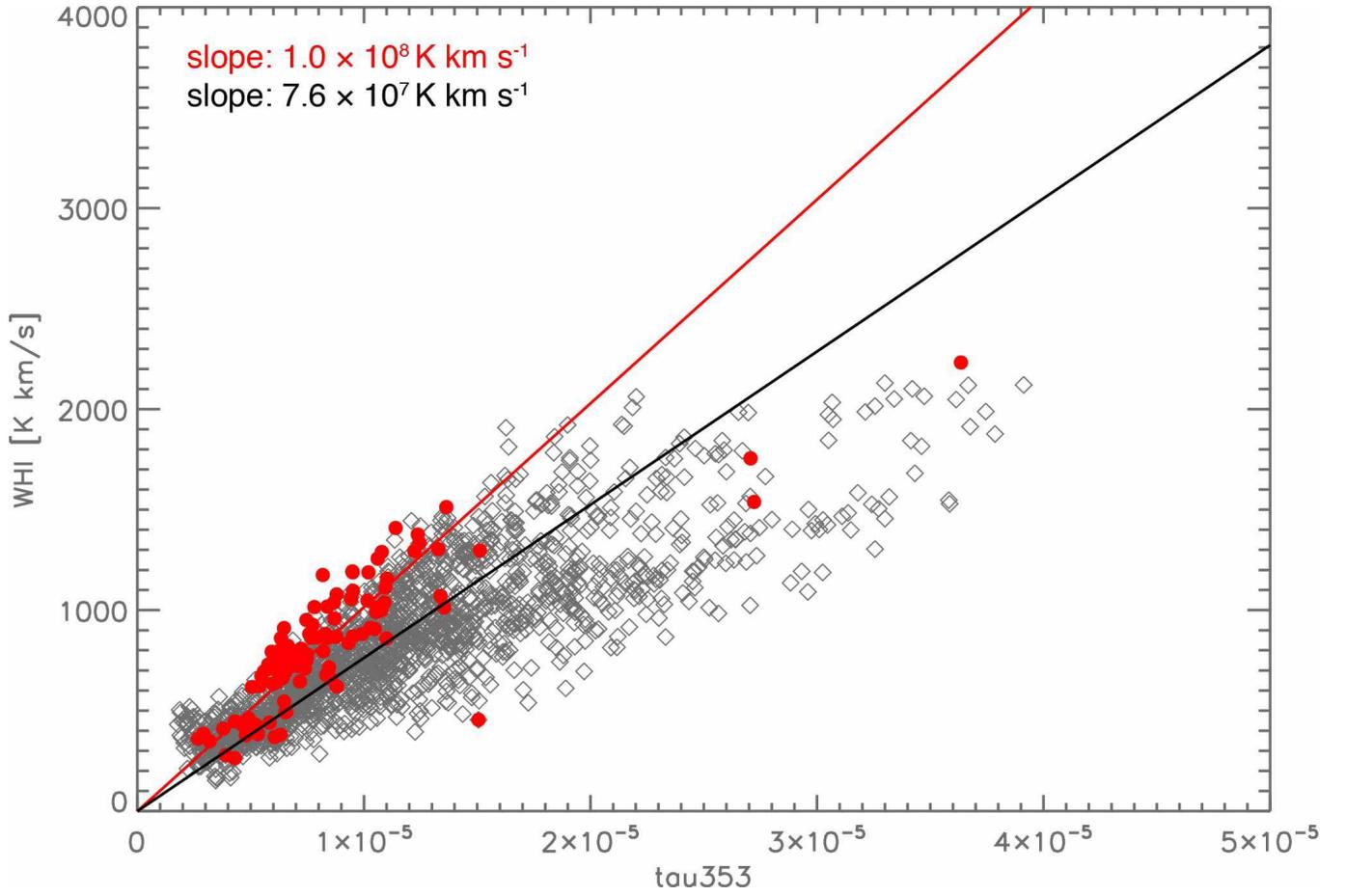}
\end{center}
\caption{Scatter plot of $W$(H{\sc i}) against $\tau$$_{353}$. The red points are the data with $T_{d}$ $\geq$ 24.5 K in the N44 region (the region of Figure 1(a)), while grey points are the data with $T_{d}$ $\geq$ 22.5 K in the Bar region (Paper I). The red and black linear lines denote the results of the least-squares fits assuming zero intercept toward the Bar region and the N44 region, respectively.}
\label{fig11}
\end{figure*}

\begin{figure*}[htbp]
\begin{center}
\includegraphics[width=\linewidth]{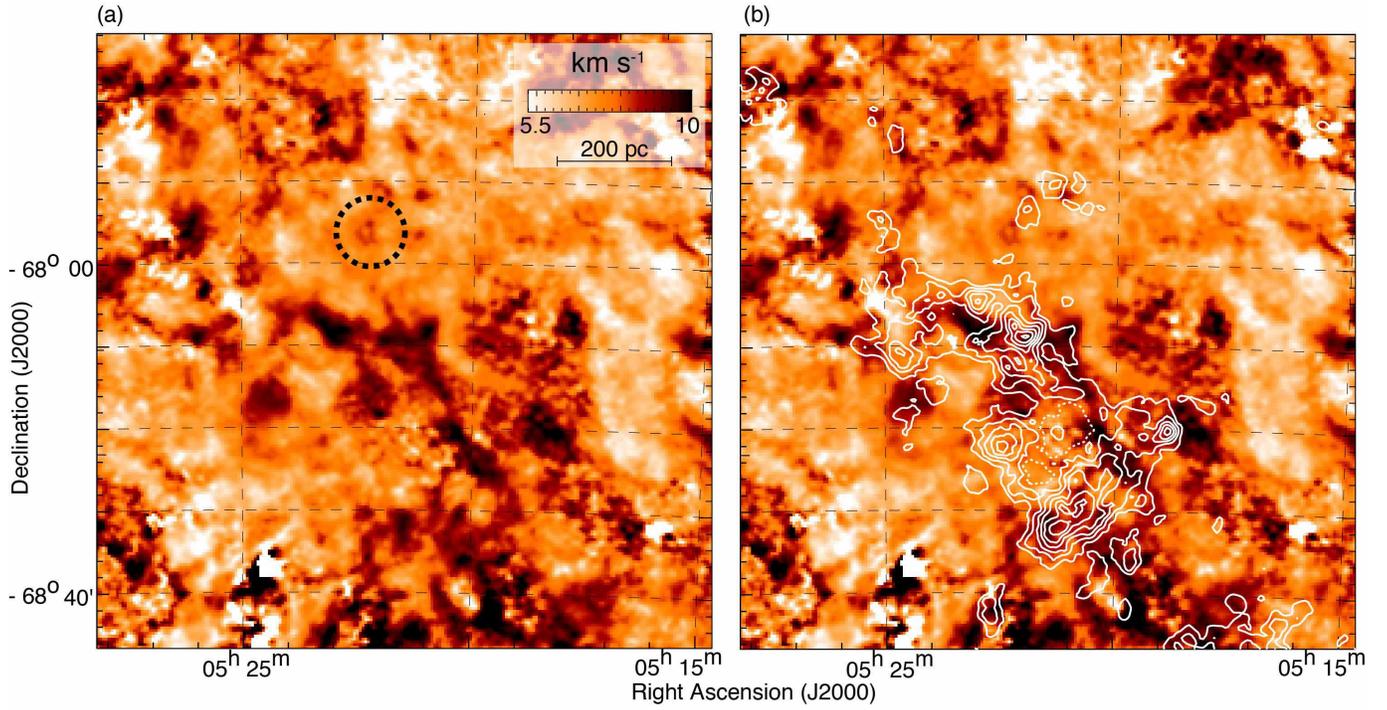}
\end{center}
\caption{(a) Second moment map of the D-component overlaid with the black dashed circle showing the location of the H$\alpha$ shell. (b) Same image as (a), but with the contours indicating the I-component. The contour levels are the same as in Figure \ref{fig5}b.}
\label{mom}
\end{figure*}

\begin{figure*}[htbp]
\begin{center}
\includegraphics[width=10cm]{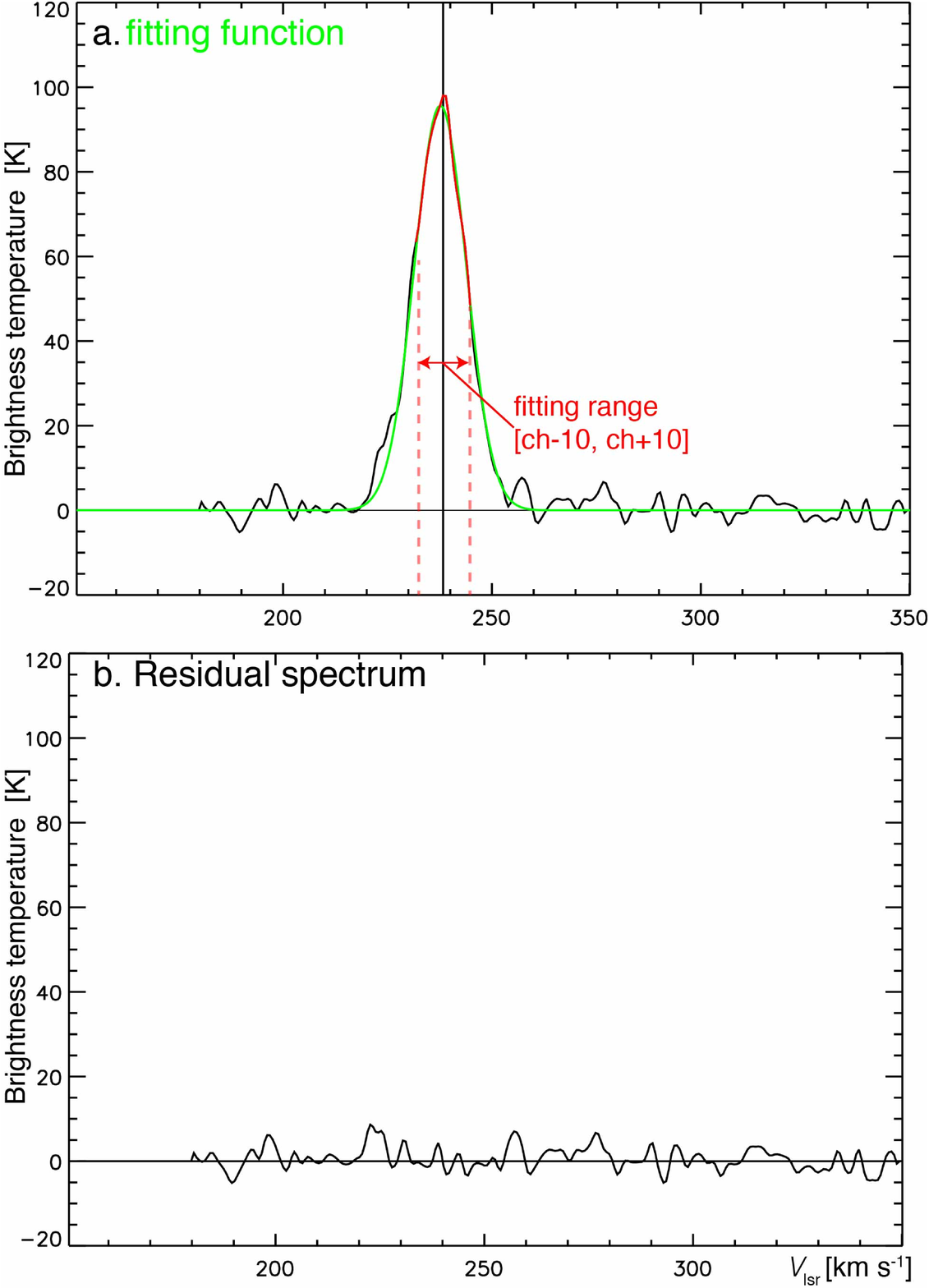}
\end{center}
\caption{The example of H{\sc i} spectral fitting. The horizontal and perpendicular axis are velocity and brightness temperature, respectively. (a) Black line indicates the original H{\sc i} profile. Green and red curved lines indicate the fitting function and H{\sc i} data used for fitting, respectively. Red dashed lines indicate the fitting range. (b) Residual spectrum of H{\sc i} spectrum.}
\label{figA1}
\end{figure*}

\begin{figure*}[htbp]
\begin{center}
\includegraphics[width=\linewidth]{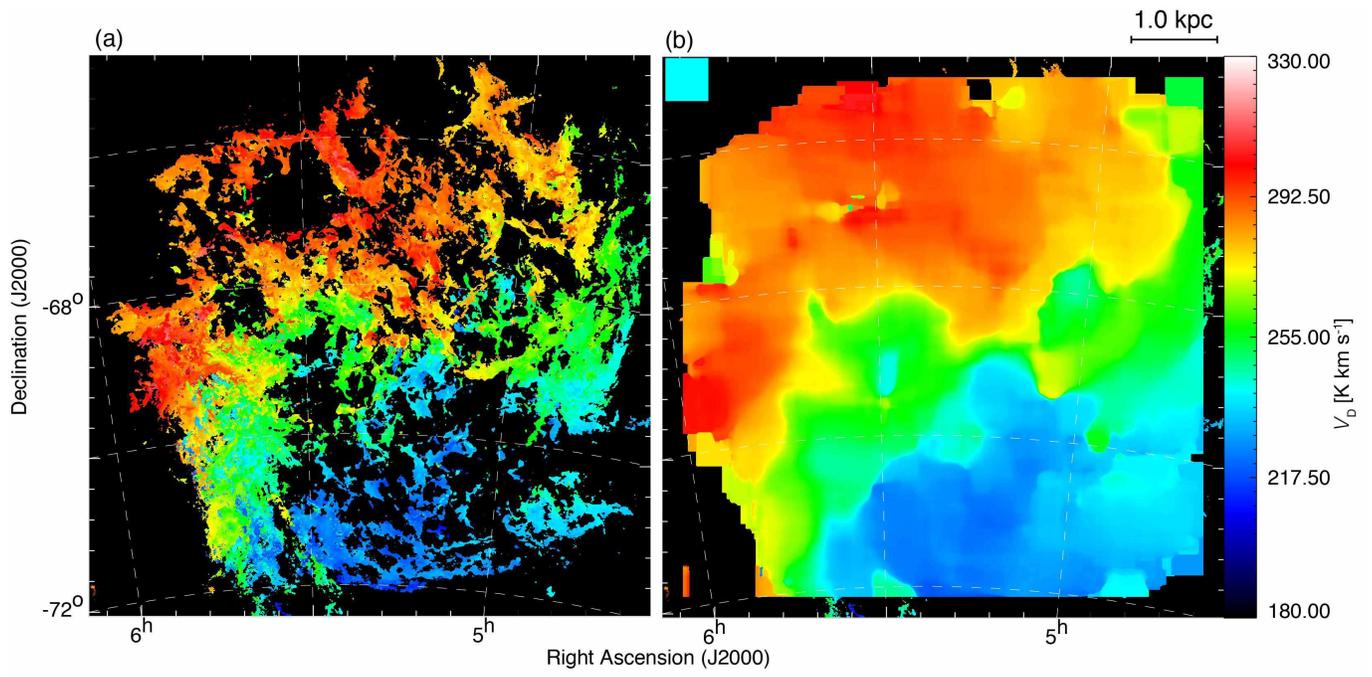}
\end{center}
\caption{(a) Radial velocity map of the D-component. Black region is masked area with weak H{\sc i} intensity. (b) Smoothed radial velocity map of the D-component. The size of median filter is also shown in the top left corner.}
\label{figA2}
\end{figure*}

Here we describe the method of deriving the rotation curve of the LMC {using} the radial velocity map of the D-component as shown in Figure 14b. 
\begin{enumerate}

\item Magellanocentric radial velocity field

First, we subtract the system velocity of the LMC from the radial velocity of the D-component for each pixel, and define $V_{\rm Magellan,ij}$ : $V_{\rm Magellan,ij}=v_{ij}-V_{\rm sys}$ where  $V_{\rm Magellan,ij}$ is Magellanocentric radial velocity of the D-component, $v_{\rm ij}$ is peak velocity of the D-component at the pixel position and $V_{\rm sys}$ is system velocity of the LMC {of 252.5 km s$^{-1}$}. The suffix of ($i,j$) denotes the coordinates along the major and minor axis in Figure \ref{figB1}b.



\item Rotation velocity field

The LMC is almost a face-on galaxy, but it has some inclination angle between the plane of the sky and the plane of the galaxy disk. So, we compensate the velocity gradient due to the inclination. We derive rotation velocity $V_{\rm rot}(R_{ij})$ at the radius of $R_{ij}$ by using equation \citep{1977ApJ...218..377W}:

\begin{eqnarray}
V_{\rm rot}(R_{ij})=\frac{V_{\rm Magellan}}{\rm cos \theta_{ij} \rm sin \it I} , 
\end{eqnarray}
where $R_{ij}$ is the radius from the center on the plane of the galaxy {given as $R_{ij}$=$\sqrt{x^{2}+(y/cos I)^{2}}$, $I$ is the inclination angle., The minor axis and major axis are drawn with the cyan lines in Figure \ref{figB1}b. $\theta_{ij}$ is the angle with respect to the major axis, and position ($i,j$).}

\item{Mask of the region}

We masked regions not used for deriving the rotation curve shown as the shaded regions in Figure \ref{figB1}b. Majority of the velocity fields follow a flat disk rotation (Figure \ref{figB1}a). However, there are some regions with velocity deviating from the galaxy rotation. This trend is also suggested by \cite{1992A&A...263...41L}. In order to minimize the local disturbances, we mask the regions with $V_{\rm rot}$ $\geq$ 120 km s$^{-1}$ or $V_{\rm rot}$ $\leq$ 0 km s$^{-1}$. In addition, a $\mid$$\theta_{ij}$$\mid$ $<$ 5$^{\circ}$-sector around the major axis is also masked, because the rotation velocity cannot be determined. 

The locally disturbed regions are roughly located at southeast arm (the H{\sc i} Ridge region) and the northwest arm-end. Their distributions are consistent with the spatial distribution of tidally interacting gas between the LMC and SMC calculated by \cite{2007PASA...24...21B} \citep[see also Figure 4 of][]{2007PASA...24...21B}. Thus, it is suspected that the local disturbances are due to the tidal interaction.

\item{Derivation of rotation curve}

We divide the whole region into 16 elliptical annulus whose widths are 0.25 kpc (white ellipses in Figure \ref{figB1}c). We plot the averaged rotation velocity ($V_{\rm rot}$) for each annulus as a function of the deprojected radial distance $R_{i,j}$ on the plane of the galaxy from the kinematic center of the LMC (red cross in Figure \ref{figB1}d). The error bar in $R_{i,j}$ is the width of the annulus and that in $V_{\rm rot}$ is give as the standard deviation of the rotation velocity in the annulus.

\item{Fitting to the flat rotation model}

We fitted the plotted data to the flat rotation model expressed as the following tanh function as following equation \citep{1997ApJ...479..244C}:
\begin{eqnarray}
V_{\rm rot}(R_{ij})=V_{\rm \infty} \rm tanh (\it R_{\rm ij}/R_{\rm 0}), 
\end{eqnarray}
where $R_{\rm 0}$ is the radius where the velocity field change from rigid to flat rotation, and $V_{\rm \infty}$ is the circular velocity at $R_{\rm ij} > R_{\rm 0}$.

There are many previous studies of the LMC kinematics as reviewed by \cite{1990A&ARv...2...29W}. Nevertheless the position angle of the major axis and the galaxy inclination are not well determined. The estimated position angles is in the range $\sim$168$^{\circ}$--220$^{\circ}$ and the inclination of 30$^{\circ}$--40$^{\circ}$. We decide the best fit values making the reduced chi-square  value given by the rotation curve fittings closest to unity by changing the values of the position angle and inclination, as listed in Table 4. 

In addition, we change the value of $V_{\rm sys}$ by 0.5 km s$^{-1}$ step between 245 km s$^{-1}$ and 260 km s$^{-1}$, and the best fit value is 252.5 km s$^{-1}$. We also optimized the kinematic center of the LMC (R.A., Dec.= 05$^{h}$17.6$^{m}$, $-$69$^{d}$2.0$^{m}$). We change the center position by 1$\arcmin$ step in a region $\pm$4$\arcmin$ from the center by \cite{1998ApJ...503..674K}, and obtain the best fit value of (R.A., Dec.) = (05$^{h}$16$^{m}$24.8$^{s}$, $-$69$^{d}$05$^{m}$58.9$^{s}$). We summarized the best fit parameters of the kinematic of the LMC in Table 5. The final rotation curve derived by this study is shown in Figure \ref{figB1}c.

\item{Median filter}

When we subtract the radial velocity of the D-component, we smooth the velocity field by median filter as describedin Appendix A, with the size of median filter optimized as follows. We change the size from 50 $\times$ 50 pc up to 950 $\times$ 950 pc, and adopt 500 $\times$ 500 pc size as the best choice in this study, based on the value of reduced chi-squared closest to 1 (Table 5).


\end{enumerate}

\clearpage


\



\begin{deluxetable}{cccccc}[b!]
\tablewidth{13 cm}
\tablecaption{Physical properties  of H{\sc i} gas\label{tab:mathmode}}
\label{tab:table1}
\tablehead{[-7pt] \multicolumn{1}{c}{Region} & ${\rm total\ mass}$   & $N$(H{\sc i})\tablenotemark{a}  & {\rm velocity\ separation} & I-component  & {\rm $\#$ \ of\ massive\ stars\tablenotemark{b}}\\
\multicolumn{1}{c}{\ }&(10$^{5} M_{\odot}$)&{\rm (10$^{21}$ cm$^{-2}$)} &{\rm (km s$^{-1}$)}&(Bridge feature) &\ }
\startdata
Line A& 8.1 & 5.6 & $\sim$60&yes&2  \\ 
Line B& 9.0 & 5.9&$\sim$20 &yes&32\\  
Line C& 8.1 & 5.7 & $\sim$60 /$\sim$20&yes&5 \\  
\enddata

\tablenotetext{a}{Peak values in each line}
\tablenotetext{b}{Total number of O-type and WR stars.}

\end{deluxetable}

\begin{deluxetable*}{ccccc}[]

\tablewidth{\linewidth}
\tablecaption{Mass of hydrogen gas}
\label{tab:table2}
\tablehead{\multicolumn{1}{c}{\ } & Mass of the & Mass of the & Mass of the & Total mass \\
 & L-comp. {[}$M_{\odot}${]} &I-comp. {[}$M_{\odot}${]}&D-comp. {[}$M_{\odot}${]}&{[}$M_{\odot}${]}}
\startdata
H{\sc i}&1.1$\times$10$^{5}$&5.3$\times$10$^{5}$&2.0$\times$10$^{6}$&3.0$\times$10$^{6}$\\
CO&1.1$\times$10$^{5}$&1.0$\times$10$^{5}$&2.4$\times$10$^{6}$ &3.0$\times$10$^{6}$\\
\enddata
\end{deluxetable*}







\clearpage

\begin{deluxetable*}{cccccccc}[]
\tablewidth{\linewidth}
\tablecaption{Comparison of N44 with H{\sc i} Ridge including R136}
\label{tab:comparison}
\tablehead{\multicolumn{1}{c}{Object} & $\#$ of & Age & Spatial extent of & fraction of & Column density & Total H{\sc i} mass \\
 & O/WR stars &{[}Myr{]}&colliding H{\sc i} gas & the SMC gas & (max) {[}10$^{22}$ cm$^{-2}${]} & {[}$M_{\odot}${]}\\
\multicolumn{1}{c}{(1)}&(2)&(3)&(4)&(5)&(6)&(7)}
\startdata
H{\sc i} Ridge & $\sim$ 400 & 1.5--4.7 & $\sim$ 1 kpc $\times$ 2.5 kpc & 0.5 & 1.0 & 1.0$\times$10$^{8}$\\
N44 & $\sim$ 40 & 5--6 & $\sim$ 800 pc $\times$ 800 pc & 0.3 & 0.6 & 2$\times$10$^{7}$\\
\enddata
\end{deluxetable*}



\clearpage


\begin{deluxetable*}{ccc|ccc}[]
\tablewidth{\linewidth}
\tablecaption{Results of the rotation curve fittings}
\label{tab:B1}
\tablehead{\multicolumn{1}{c}{P.A.} & Inclination & $\chi^{2}$/d.o.f & P.A.  & Inclination & $\chi^{2}$/d.o.f \\
{[}deg.{]}&{[}deg.{]}& &{[}deg.{]}&{[}deg.{]}& \\
\multicolumn{1}{c}{(1)} & (2) & &(3)&(4)& }
\startdata
\multirow{7}{*}{190} & 38          & 1.279   & \multirow{7}{*}{200} & 38          & 0.965   \\  
                     & 37          & 1.259   &                      & 37          & 0.967  \\ 
                     & 36          & 1.246   &                      & 36          & 0.978  \\  
                     & 35          & 1.248   &                      & 35          & \textbf{1.007}   \\ 
                     & 34          & 1.259   &                      & 34          & 1.054   \\
                     & 33          & 1.275   &                      & 33          & 1.146   \\
                     & 32          & 1.296    &                      & 32          & 1.271   \\ \hline
\multirow{7}{*}{195} & 38          & 1.208   & \multirow{7}{*}{205} & 38          & 0.844   \\
                     & 37          & 1.189  &                      & 37          & 0.865  \\  
                     & 36          & 1.181   &                      & 36          & 0.905  \\ 
                     & 35          & 1.190   &                      & 35          & 0.965  \\  
                     & 34          & 1.218   &                      & 34          & 1.054   \\ 
                     & 33          & 1.261    &                      & 33          & 1.162   \\  
                     & 32          & 1.329   &                      & 32          & 1.307   \\ \hline
\enddata
\tablecomments{Cols. (1)(3): Position Angle of major axis. Cols. (2)(4): Inclination of the rotation axis to the line of sight.}
\end{deluxetable*}


\begin{deluxetable*}{ll}[]
\tablewidth{\linewidth}
\tablecaption{The best fit kinematic parameters of the rotation curve}
\label{tab:B2}
\tablehead{\multicolumn{1}{l}{Parameters} & values} 
\startdata
Kinematic center (R.A., Dec.) & 05$^{\mathrm{h}}$16$^{\mathrm{m}}$24.82$^{\mathrm{s}}$, $-$69$^{\circ}$05$\arcmin$58.88$\arcsec$ \\
P.A. of majojr axis [deg.]                                                                     & 200                                          \\ 
Inclination [deg.]                                                                             & 35                                        \\ 
$V_{\infty}$ [km s$^{-1}$]& 61.1\\
$R_{\rm 0}$ [kpc]& 0.94\\
$V_{sys}$ [km s$^{-1}$]                                                                                    & 252.5                             \\ 
\enddata
\end{deluxetable*}


\begin{deluxetable*}{cc}[]
\label{tab:B3}
\tablewidth{\linewidth}
\tablecaption{$\chi$$^{2}$ fitting results with different size of the median filter}
\tablehead{\multicolumn{1}{c}{Filter size} & $\chi^{2}$/d.o.f }
\startdata
50 $\times$ 50  pc                 & 1.518          \\ 
200 $\times$ 200  pc               & 1.267         \\ 
350 $\times$ 350  pc                & 0.974         \\ 
500 $\times$ 500  pc               & \textbf{1.007} \\ 
650 $\times$ 650  pc               & 1.039           \\ 
800 $\times$ 800  pc               & 0.971         \\ 
950 $\times$ 950  pc               & 0.743           \\ 
\enddata
\end{deluxetable*}



\end{document}